\documentclass[referee]{aa}  
\usepackage{graphicx}
\usepackage{txfonts}
\usepackage{hyperref}
\usepackage[squaren,Gray]{SIunits}
\usepackage{numprint}
\usepackage{grffile}
\usepackage{color}

\DeclareRobustCommand{\rchi}{{\mathpalette\irchi\relax}}
\newcommand{\irchi}[2]{\raisebox{\depth}{$#1\chi$}} % inner command, used by \rchi

\begin{document}

\title{Frequency resolved radio and high-energy emission of pulsars}
%\subtitle{}
\author{Q. Giraud
	%\inst{1}
	\and
	J. P\'etri%\fnmsep\thanks{Just to show the usage of the elements in the author field}
}

\titlerunning{Frequency resolved emission of pulsars}

\institute{Universit\'e de Strasbourg, CNRS, Observatoire astronomique de Strasbourg, UMR 7550, F-67000 Strasbourg, France.\\
	\email{quentin.giraud@astro.unistra.fr}
}

\date{Received; accepted }

% \abstract{}{}{}{}{} 
% 5 {} token are mandatory

\abstract
{Pulsars are detected as broadband electromagnetic emitters from the radio wavelength up to high and very high energy in the MeV/GeV and sometimes even in the TeV range. Multi-wavelength phase-resolved spectra and light curves offer an unrivalled opportunity to understand their underlying radiation mechanisms and to localize their emission sites and therefore the particle acceleration regions.}
{In this paper, we compute pulsars multi-wavelength phase-resolved light-curves and spectra assuming that curvature radiation operates from inside the magnetosphere of a rotating vacuum magnet. Radio emission arises from the polar caps whereas gamma-ray energy emanates from the slot gaps in the vicinity of the separatrix between closed and open field lines.}
{By integrating particle trajectories within the slot gaps, we compute energy dependent photon sky maps in the radio band (MHz-GHz) and in the gamma-ray band (MeV/GeV) for mono-energetic %as well as for power law 
	distribution functions of leptons.}
{We obtained many details of the energy dependent light curves and phase-resolved spectra from the radio wavelength up to the gamma-ray energies. Choosing Lorentz factors of $\gamma \approx 30$ for radio emitting particles and $\gamma \approx 10^7$ for gamma-ray emitting particles, we found realistic spectra accounting for the wealth of multi-wavelength pulsar observations.}
{}


\maketitle

\section{Introduction}

Neutron stars are mainly observed as pulsars, emitting regular pulses in the whole electromagnetic spectrum from radio, through optical, X-rays up to gamma-rays. Although it is well established that the pulsar surrounding must be populated with electron-positron pairs, the localisation and dynamics of efficient particle acceleration and radiation sites are still ill understood. Nevertheless, several attempts to design such regions at least geometrically have been proposed like the outer gaps \citep{cheng_energetic_1986}, the slot gaps \citep{arons_pair_1983} and the polar caps \citep{ruderman_theory_1975}. By computing light-curves from these locations, it is then possible to compare the merit of each site and test their ability to reproduce the observed light-curves \citep{dyks_two-pole_2003, dyks_relativistic_2004}. The presence of a plasma partially or completely screening the electric field shows up in distortions of the light-curves from a vacuum rotator compared to a force-free model \citep{bai_modeling_2010}. Often in the vacuum field investigations, widely used in the literature, the associated accelerating electric fields are not taken into account self-consistently. Let us however mention the work of \cite{kalapotharakos_gamma-ray_2012} and \cite{kalapotharakos_gamma-ray_2014}, who indeed used resistive plasma models with low conductivity to mimic almost vacuum electromagnetic fields. They also produced sky-maps and light-curves taking into account the accelerating electric field.

\cite{watters_atlas_2009} compiled an atlas of geometric light curves for young pulsars showing the essential characteristics of gamma-ray pulse profiles depending on viewing angle and obliquity. \cite{romani_constraining_2010} then designed a tool to constrain the magnetospheric structure from these gamma-ray light curves. \cite{venter_probing_2009} investigated the special population of millisecond gamma-ray pulsars showing that two-pole caustics and outer gap models are favoured. See also \citet{pierbattista_light-curve_2015, pierbattista_young_2016} for a large sample of pulsars fitted with several emission models and \citet{johnson_constraints_2014} for a similar study about millisecond pulsars. \citet{harding_gamma-ray_2011}  produced atlases of two-pole caustics and outer gap emission models in a force-free and vacuum retarded dipole field geometry to compare light curve features in symmetric and asymmetric slot gap cavities. Obviously, more constraints can be obtained from simultaneous radio and gamma-ray fitting, see for instance the work by \cite{petri_unified_2011} assuming gamma-ray emission from the striped wind and radio emission from the polar caps.

The second Fermi gamma-ray pulsar catalogue \citep{abdo_second_2013} contains plenty of information about gamma-ray pulsar spectra and light-curves. The gamma-ray peak separation clusters around $\Delta \approx 0.5$ and the radio peak usually leads the first gamma-ray peak but with some outliers. Force-free or ideal MHD computations are unable to self-consistently accelerate particles and localize the emission site. Some kind of dissipation of the electromagnetic field is required in order to produce a signal detectable on earth. So dissipation within the magnetosphere and/or wind must occur, but the precise mechanism and its efficiency are difficult to predict from first principles. Nevertheless, some dissipative magnetospheres, called force-free-inside-dissipative-outside (FIDO) and introduced by \cite{kalapotharakos_toward_2012}, were used by \citet{brambilla_testing_2015} for computing the phase-averaged and phase-resolved $\gamma$-ray spectra of eight of the brightest Fermi pulsars. They used billions of test particles trajectories to compute curvature radiation spectra in realistic fields of $\numprint{e7}-\numprint{e9}$~T. Based on this work \cite{kalapotharakos_fermi_2017} constrained the dissipation mechanism by looking at curvature radiation in the equatorial current sheet outside the light-cylinder, using Fermi/LAT spectral data. This could put limits on the strength of the accelerating electric field. They used test particle integration in the radiation reaction limit regime in the global force-free dissipative magnetosphere which is basically a fluid description avoiding the stringent strong field constrain faced by PIC codes. In such a way, they were able to deduce spectra for realistic pulsar field strengths and periods. Nevertheless, starting from PIC simulations, \cite{kalapotharakos_three-dimensional_2018} recently found a relation between the particle injection rate and the spin down luminosity. This work shows the fruitful feedback between simulations and observations to extract useful information about the nature of particle acceleration and dissipation of the relativistic magnetized flow. \cite{harding_gamma-ray_2016} and \cite{venter_high-energy_2018} gave recent reviews of the successful interplay between magnetospheric modelling and gamma-ray observations.

Further attempts to fit particular pulsars were carried out by other groups. For instance \cite{takata_polarization_2007} and \cite{hirotani_outer-gap_2008} used the vacuum retarded dipole to model the outer gap of the Crab pulsar. \citet{du_gamma-ray_2011} performed computation in the annular gap context for the Vela pulsar whereas \citet{du_radio--tev_2012} did it for the Crab pulsar. Several millisecond pulsars were also fitted by \citet{du_radio_2013} using a static dipole.
Recently the impact of general-relativistic effects on pulsar light-curves have been investigated by \cite{petri_general-relativistic_2018} followed by some improvement made by \cite{giraud_radio_2020}, including the Shapiro delay.

In these aforementioned works, our attention was focused on the shape of the pulses emitted in radio and gamma-ray, explicitly showing the impact of gravitation on the geometry of the profiles and on the phase shift between the main radio pulse and the first gamma-ray pulse. We did not take into account the frequency characteristics of this emission, assuming it to be curvature radiation, which is a function of the point of emission and more particularly of the curvature of the field lines. But what about their frequency distribution, i.e. the radio and gamma-ray spectrum? How does the shape of the pulses evolve as a function of frequency or energy? This paper answers these questions by providing precise quantitative results if the emission comes from curvature radiation all along the magnetic field lines in a flat spacetime geometry. The outline of our paper is as follows.
In section~\ref{sec:Model} we expose our model magnetosphere model and the associated radiation mechanism. Next, in section~\ref{sec:HE}, we show detailed results of high-energy spectra and light-curves from the slot gap models, localising the sites of efficient gamma-ray photon production depending on the magnetic inclination angle and line of sight. In section~\ref{sec:radio}, the same approach is used to study the radio spectra and light-curves depending on the geometry. Conclusions and possible extensions are summarised in section~\ref{sec:Conclusion}.

\section{Magnetospheric emission model
\label{sec:Model}}

\subsection{A coherent multi-wavelength model}

Multi-wavelength modelling of radio-loud gamma-ray pulsars is usually limited to the joint computation of only one radio and one gamma-ray light curve. Although this strategy already makes it possible to constrain the geometrical parameters of the magnetosphere, this is hardly satisfactory because the pulse profiles evolve as a function of frequency or energy. A detailed multi-frequency investigation, including coherently the calculation of radio and gamma-ray spectra from, for example, the curvature radiation, proves to be much more constraining and will allow to physically and quantitatively link the geometry of the field lines, the altitude of the photon production sites, the distribution functions of the emitting particles as well as the associated light curves.

In this paper, this detailed view of broadband electromagnetic emission processes is discussed in depth. To do so, we start with a reminder of the method of calculation of the properties of the curvature emission, its power and its characteristic frequency, applied to the pulsar magnetosphere. Then we will determine the high energy spectrum around the GeV, thus in the Fermi/LAT band, as well as some characteristic light curves to conclude on the spectrum and radio pulse profiles.

\subsection{Magnetosphere structure}

We use the analytical expression of the electromagnetic field of a rotating dipole in vacuum as first given by \cite{deutsch_electromagnetic_1955}. Charged particles are accelerated in gaps where the plasma present in the pulsar's magnetosphere is not dense enough to screen the component of the electric field parallel to the magnetic field lines. In our model, those gaps are either located along the last closed magnetic field lines from the star surface up to the light cylinder (slot gaps, \citep{arons_pair_1983}) or above the polar caps delimited by the last closed magnetic field lines (polar gaps, \citep{ruderman_theory_1975}). We also assume that radio and gamma-ray emission are produced by the acceleration of charged particles in those gaps through curvature radiation. Let us therefore remind the essential feature of this emission mechanism.

\subsection{Curvature radiation}

The calculation of the frequency of curvature radiation is deduced from the radius of curvature~$\rho$ of the particle trajectory along the magnetic field lines, which is equal to the radius of curvature of these field lines in the corotating frame. The characteristic curvature frequency~$\omega_{\rm curv}$ is \citep{jackson_electrodynamique_2001}
\begin{equation}
\omega_{\rm curv} = \dfrac{3}{2}\gamma^{3}\dfrac{c}{\rho}
\label{eq:frequence_courbure}
\end{equation}
beyond which the particle no longer radiates significantly. Specifically, the shape of the curvature radiation spectrum for a particle with Lorentz factor~$\gamma$ is given in \cite{jackson_electrodynamique_2001} by
\begin{equation}\label{eq:spectre_courbure}
\frac{dI}{d\omega} = \frac{\sqrt{3}}{4\,\pi\,\varepsilon_0} \, \frac{e^2}{c} \, \gamma \, F\left(\frac{\omega}{\omega_{\rm c}}\right) \quad ; \quad F(x) = x \, \int_{x}^{+\infty} K_{5/3}(t) \, dt
\end{equation}
where $K_{5/3}$ is the modified Bessel function of order~$5/3$ \citep{arfken_mathematical_2005}. 
%A very good approximation of the function is according to \cite{aharonian_angular_2010}
%\begin{equation}\label{eq:spectre_courbure_approximation}
%F(x) = 2.15 \, x^{1/3} \, (1+3.06\,x)^{1/6} \, \frac{1 + 0.884\,x^{2/3} + 0.471 \, x^{4/3}}{1 + 1.64\,x^{2/3} + 0.974 \, x^{4/3}} \, e^{-x} .
%\end{equation}
%The shape of the spectrum is shown in figure~\ref{fig:Spectre_Courbure} in red line and the approximation in blue line. Since the accuracy is better than 0.2\% over the entire range, the curves cannot be distinguished with the naked eye.
%\begin{figure}[tbph]
%	\centering
%	\includegraphics[width=\columnwidth]{spectre_courbure.eps}
%	\caption{Spectra of curvature radiation around the characteristic frequency~$\omega_{\rm curv}$. The approximation in blue is undistinguishable from the exact expression in red.}
%	\label{fig:Spectre_Courbure}
%\end{figure}

A trajectory or line in space has a curvature as well as a twist. Locally, at a point of this curvature, we associate a trihedron $(\vec{T}, \vec{N}, \vec{B})$ also called Frenet's frame. The curvature~$\kappa$ indicates the change in the tangent vector~$\vec{T}$ and the torsion~$\tau$ indicates the variation in the osculating plane at the same point. In summary the Frenet formulas giving the variations of the basis vectors of the trihedron are
\begin{subequations}
	\begin{align}
	\frac{d\vec T}{ds} & = \kappa \, \vec N \\
	\frac{d\vec N}{ds} & = - \kappa \, \vec T + \tau \, \vec B \\
	\frac{d\vec B}{ds} & = - \tau \, \vec N
	\end{align}
\end{subequations}
$ds$ being the curvilinear abscissa along the trajectory.
Torsion does not intervene in the calculation of the particle curvature radiation. The curvature itself is thus deduced from the derivative of the tangent $\vec{T} = \vec{B}/B$ to the field lines as a function of the curvilinear abscissa~$s$ along these field lines. The curvature radius $\rho$ is the inverse of the curvature~$\kappa$ of the magnetic field which is the derivative of the tangent $\mathbf{T}$ to the magnetic field line with respect to the curvilinear abscissa~$ds$, so we have
\begin{equation}
\kappa = \dfrac{1}{\rho} = \left\|\frac{d\vec{T}(s)}{ds}\right\| .
\label{eq:courbure}
\end{equation}
We recall that the variation of the curvilinear abscissa~$ds$ is related to the variations of the coordinates of the trajectory $dx^i$ by $ds^2 = \gamma_{ik} \, dx^i \, dx^k$ in any spatial geometry given by the spatial metric~$\gamma_{ik}$ which is the spatial projection of the space-time metric~$g_{ik}$ with respect to a given observer. In Minkowski's space-time in Cartesian coordinates, this expression simplifies to $dx^i = (dx,dy,dz)$ and $ds^2 = dx^2+dy^2+dz^2$. 

In our model, we assume that the particles follow the field lines in the reference frame corotating with the star, the frequency of the curvature radiation~$\omega_{\rm c}'$ is therefore that of the rotating frame of reference \eqref{eq:frequence_courbure}. In order to find the frequency of the radiation emitted by our rotating pulsar and measured by an distant inertial observer at rest $\omega_{\rm c}$, we must take into account the Doppler effect and perform a Lorentz transformation from the corotation frame to the inertial frame of the observer. This is equivalent to multiplying the equation~\eqref{eq:frequence_courbure} by a Doppler factor~$\eta$ due to the neutron star rotation. We thus have $\omega_{\rm c} = \eta \, \omega_{\rm c}'$ with the Doppler factor generally put in the form
\begin{equation}
\eta = \dfrac{1}{\gamma\,(1-\vec{\beta}\cdot{\vec{n}})}
\label{eq:facteur_doppler}
\end{equation}
$\vec{n}$ being the unit vector giving the initial direction of propagation of the photon (tangent to the magnetic field line) as seen in the observer's frame of reference and taking into account the aberration and $\vec{\beta}$ the corotation velocity. So $\vec{n}$ is given by the equation
\begin{equation}
\vec{n} = \frac{1}{\eta}\left[ \vec{n}'+\gamma\left( \dfrac{\gamma}{\gamma+1}(\vec{\beta} \cdot \vec{n}')+1\right) \vec{\beta}\right] .
\label{aberration_restreinte}
\end{equation}

$\gamma$ is the Lorentz factor of the particle accelerated in the electromagnetic field, primary or secondary beam. 
The power radiated by this particle is given by
\begin{equation}
P_{\rm curv} = \dfrac{e^{2}\,\gamma^{4}\,c}{6\,\pi\,\epsilon_{0}\,\rho^{2}}
\label{eq:power}
\end{equation}
with $e$ the elementary charge and $\epsilon_0$ the vacuum permittivity. 

For curves of known geometry, this curvature can be determined analytically, as for example for a static dipole for which the equation of the field lines is known, the lines being themselves each contained in a plane. For a rotating dipole such as the Deutsch dipole or in general relativity, it is not possible to obtain an analytical expression of this curve. It must be estimated numerically by approximating the derivative~\eqref{eq:courbure}. 

Concretely, this curvature is calculated by measuring the variation $\Delta \vec{T}$ of the vector tangent $\vec{T}$ to the magnetic field lines as one moves along them by a value $\Delta s$ of the curvilinear abscissa $s$ small enough that the finite difference represents a good approximation of the derivative. Typically a fraction of the radius of the star $\Delta s \lesssim R_*$ is sufficient. We then replace $\kappa$ in the equation~\eqref{eq:courbure} by a discretized version.
We calculate this derivative vector field by measuring the variation of the tangent vector~$\mathbf{T}$ along a field line by varying the curvilinear abscissa~$s$
\begin{equation}\label{eq:tangente}
\frac{d\mathbf{T}}{ds} %\approx \frac{\Delta\vec{T}}{\Delta s} 
\approx \frac{\vec{T}(s+\Delta s/2)-\vec{T}(s-\Delta s/2)}{\Delta s}
\end{equation}
%\begin{equation}
%\kappa \approx %\left\| \dfrac{\Delta \vec{T}}{\Delta s}\right\| =
%\dfrac{ \left\| \Delta \vec{T}\right\| }{\Delta s}.
%\end{equation}
In order to minimize discretization errors, we chose a method with centered finite differences.
% such that the variation of the tangent vector depending on the curvilinear abscissa is :
%\begin{equation}
%\Delta\vec{T} \approx \vec{T}(s+\Delta s/2)-\vec{T}(s-\Delta s/2) + o(\Delta s^2).
%\end{equation}
Knowing the local radius of curvature and the value of the magnetic field at the point of emission in the rotating frame of reference, we switch back to the spectrum of the curvature radiation in the inertial frame of reference by imposing an energy distribution function for the photon emitting leptons. The simplest situation consists in taking a mono-energy distribution of particles although a general particle population is possible.%, the spectrum of which has been recalled in chapter~\ref{chap:intro}.

\subsection{Particle population}

The shape of the light curves and spectra strongly depend on the energy distribution of the radiating particles. Current models of acceleration of primary particles and creation of secondary and higher generation of pairs (tertiary, quaternary, ...) predict a peak of the Lorentz factor around $\gamma\approx10^7$ for primaries and around $\gamma\approx10^2$ for others \citep{beskin_physics_1993}. Although our approach is easily adapted to any particle distribution function, in this work we only consider mono-energetic particles either in the primary beam or in the secondary beam. The primary particles are responsible for the high-energy emission in the GeV band in the slot gap while the secondary particles from the cascades produce the radio emission from the polar caps.

In a more realistic picture, the particle distribution function significantly deviates from a mono-energetic one, the radiation being linearly additive, the calculation of the spectra and light curves will follow the same linear addition scheme of the individual mono-energetic contributions with respective weights corresponding to the weights imposed by the particle distribution function. For example, for a population containing $N_1$ particles of energy $\gamma_1$ and $N_2$ particles of energy $\gamma_2$, the total emission will be $N_1$ times that of the mono-energy spectrum at $\gamma_1$ to which we superimpose $N_2$ times that of the mono-energy spectrum at $\gamma_2$. A continuous distribution is then sub-divided into several intervals of mono-energetic particles and the entire spectrum can be reconstructed. Our results therefore serve as a building block for the most general distribution function obtained by a post-treatment of mono-energetic distribution functions.

Let us now review the results for the high-energy mono-energetic and radio spectra: for each of the two cases, we will assume that at each emission point, located along the last magnetic field lines with a curvilinear spacing~$\Delta \ell$ between each point, there is emission by curvature radiation considered to emanate from particles with Lorentz factors $\gamma$ fixed for each wavelength. We assume that there is no emission for points located more than $95\%$ of $R_{cyl}$ from the centre of the star. 

The classical estimation of the maximum Lorentz factor attainable by particles in the electromagnetic field takes into account their braking by radiation reaction in order to compensate the acceleration by the parallel electric field $E_\parallel$ present in the gaps. The precise value of this field is strongly dependent on the dynamics within the gaps. By equating the radiated power \eqref{eq:power} and the power provided by this electric field $e\,E_\parallel\,c$, the Lorentz factor becomes
\begin{equation}\label{eq:gamma_max}
\gamma_{\rm max} = \left( 6\,\pi\,\varepsilon_0 \, \frac{E_\parallel \, \rho^2}{e} \right)^{1/4} \approx 10^7. 
\end{equation}
In our vacuum model, $E_\parallel \approx 10^{12}$~V/m corresponds to the upper limit, that of the Deutsch field where the electric field is not screened at all. In reality, this value is much lower but the maximum Lorentz factor only varies with $E_\parallel^{1/4}$, so it is not very sensitive to a large variation of this field.

It is currently well accepted that the pulsar magnetosphere contains at least two different populations of particles flowing out along open magnetic field lines. The first component is made of primary particles accelerated by the strong parallel electric field in vacuum gaps up to Lorentz factors about $\gamma \approx \numprint{e7}$ and with a particle density number close to the corotation density of $n_p \approx 2 \, \varepsilon_0 \, \Omega \cdot B / e$. These primary particles emit gamma-ray photons decaying into a secondary flow made of electron-positron pairs \citep{erber_high-energy_1966} with Lorentz factor in the range $\gamma \approx \numprint{e2}-\numprint{e3}$ and a multiplicity factor of~$\kappa \approx \numprint{e3}-\numprint{e5}$ \citep{daugherty_electromagnetic_1982}. The typical energy of curvature photons produced by each beam will therefore vary by several orders of magnitude. The radio emission is generated by the dense secondary beam whereas the high energy MeV/GeV emission is generated by the ultra-relativistic but much less dense primary beam.

In our model, we assumed a simple mono-energetic population, 
%In our model, we tried two particle distribution functions, a simple mono-energetic population and a power-law distribution function for both the primary particles and the secondary pairs. They are given respectively by $n(\gamma) \propto \delta(\gamma - \gamma_0)$ and by $n(\gamma) \propto \gamma^{-p}$ for $\gamma \in [\gamma_{\rm min}, \gamma_{\rm max}]$ where $p$ is the power-law index of the particle energy spectrum and $\gamma_{\rm min}, \gamma_{\rm max}$ the minimum and maximum Lorentz factor. For the primary flow, 
we typically set $\gamma_{\rm g} = \numprint{e7}$ for the gamma-rays emission coming from the slot gaps whereas for the radio emission from the polar caps we take $\gamma_{\rm r} = 30$.% see table~\ref{tab:beam}.

%\begin{table}
%\begin{center}
%\begin{tabular}{|cccc|}
%	\hline 
%	Energy band & $\gamma_0$ & $\gamma_{\rm min}$ & $\gamma_{\rm max}$ \\
%	\hline 
%	Radio	& \numprint{e2} & & \\ 
%	Gamma-ray & \numprint{e7} & & \\ 
%	\hline 
%\end{tabular} 
%\caption{Typical parameters for the primary and secondary beam. \cor{Peux-tu compléter et vérifier les valeurs for les différents $\gamma$?} \label{tab:beam}}
%\end{center}
%\end{table}

\subsection{Emissivity}

First of all, the calculation of the spectra we present in the following does not take into account the range of frequencies at which photons are emitted. We approximate the spectrum by a Dirac distribution centred on the characteristic frequency $\omega_c$ and whose intensity corresponds to the total power of the integrated radiation on all frequencies of the true continuous spectrum \eqref{eq:power}. In other words, the spectrum is approximated in the corotating frame by
\begin{equation}\label{eq:spectre_sync_mono}
\frac{dI'_{\rm curv}}{d\omega'} = P'_{\rm curv} \, \delta(\omega' - \omega'_{\rm c}).
\end{equation}
Each point on a field line emits photons at a single frequency given by the local characteristic frequency~$\omega'_{\rm c}(\rho)$ \eqref{eq:frequence_courbure} associated with the local curvature of this same field line. The number of photons to consider is proportional to the total emitted power~$\omega'_{\rm curv}(\rho)$. A Lorentz transform bring all these quantities into the observer inertial frame. This contrasts sharply with our previous work \citep{giraud_radio_2020} where each position on a given field line produced only one photon of indeterminate energy. In this paper, we remove the ambiguity on frequency by incorporating details of the emission mechanism, frequency and power. Emissivity will no longer be constant along a field line but will vary according to the local curvature.

In a forthcoming step, we will consider using the full expression of the curvature radiation spectrum given by \eqref{eq:spectre_courbure}. Finally, in another possible extension, the most realistic situation takes into account a particle distribution function depicted by a power law such that the number of particles of Lorentz factor between $\gamma$ and $\gamma+d\gamma$ is given by
\begin{equation}\label{eq:distribution_function}
\frac{dN}{d\gamma} \propto \gamma^{-p} \qquad \textrm{ with } \qquad \gamma \in [\gamma_{\rm min}, \gamma_{\rm max}]
\end{equation}
where $p$ is the spectral index of the power law, the minimum Lorentz factor being~$\gamma_{\rm min}$ and a the maximum value being~$\gamma_{\rm max}$. This particle distribution function produces another power law for the curvature emissivity~$j_{\rm curv}$ such that
\begin{equation}\label{eq:emissivite}
j_{\rm curv} = \int \gamma^{-p} \, \frac{dI_{\rm curv}}{d\omega} \, d\gamma \propto \omega^{-(p-2)/3}.
\end{equation}
There is thus a simple relationship between the spectral index of the power distribution of particles and photons. Beyond the cut-off frequency associated with $\gamma_{\rm max}$ the emissivity falls exponentially according to $e^{-\omega/\omega_{\rm c}^{\rm max}}$. Below the cut-off frequency associated with $\gamma_{\rm min}$ the emissivity decreases according to another power law independent of the particle distribution and scaling as~$(\omega/\omega_{\rm c}^{\rm min})^{1/3}$. In between these two limits, the photon spectral index is related to particle power-law distribution~$p$.

\subsection{Aberration, retardation and seep-back effects \label{sec:retard}}

The radio and high-energy emission of pulsars originate at high altitude in the magnetosphere, and well above the surface for young pulsars of period~$P\gtrsim100$~ms \citep{mitra_nature_2017}. The corotation of these emission sites at relativistic velocities imparts unique characteristics to the pulsar profiles by impacting the relationship between the geometry of the emitting zones and their observational signature.

\cite{phillips_radio_1992} has detailed the impact of several important effects on the arrival time of pulses of all kinds. Let us consider two emission zones located at an altitude respectively~$r_1$ and $r_2>r_1$. For photons moving radially away from the neutron star, the delay introduced by the difference in path to be travelled becomes a phase shift for the observer
\begin{equation}\label{eq:decal_ret_mink}
\Omega \, \Delta t_r = \frac{r_1-r_2}{R_{\rm cyl}}.
\end{equation}
The dragging of the photon production areas by the rotation of the pulsar causes a projection of the direction of propagation of the photons in the direction of rotation due to aberration by a value of
\begin{equation}\label{eq:aberration_angle}
\theta_a = \arctan \left(\frac{v_\phi}{c}\right) = \arctan \left( \frac{r\,\sin\rchi}{R_{\rm cyl}} \right).
\end{equation}
For both emission sites, this introduces an additional time delay of
\begin{equation}\label{eq:decal_aberr}
\Omega \, \Delta t_a =  \arctan \left( \frac{r_1\,\sin\rchi}{R_{\rm cyl}} \right) - \arctan\left(  \frac{r_2\,\sin\rchi}{R_{\rm cyl}} \right) .
\end{equation}
Finally, the rotation of the dipole curves the field lines in the opposite direction of rotation. \cite{shitov_period_1983} gave a simple expression for this sweep-back effect, such that the delay induced amounts to
\begin{equation}\label{eq:retard_shitov}
\Omega \, \Delta t_B = 1.2\,\sin^2\rchi \, \left[ \left( \frac{r_2}{R_{\rm cyl}} \right)^3 - \left(  \frac{r_1}{R_{\rm cyl}} \right)^3 \right] .
\end{equation}
The total delay~$\Delta t_\Sigma$ is then the sum of each of these contributions $\Delta t_\Sigma = \Delta t_a + \Delta t_r + \Delta t_B$.

We now discuss in some details the high-energy and radio light-curves and spectra for mono-energetic particle distribution functions applied to the polar cap and slot gap models of a vacuum rotating dipole.

\section{High-energy emission\label{sec:HE}}

We are interested in the high-energy emission as measured in the Fermi/LAT (Large Area Telescope) band, between 100~MeV and 100~GeV. This Fermi/LAT telescope has detected nearly 300~gamma-ray pulsars all showing very similar spectra peaking around several GeV \citep{abdo_second_2013}. Our long lasting aim is to compare our predicted light curves and spectral modelling with a whole population of young and millisecond radio-loud gamma-ray pulsars. This section details the bottom line to construct general light-curves and spectra  from the slot gaps knowing the particle distribution function.

\subsection{Thin slot gap}

The locus of the last open field lines taking their root at the polar caps generates a two-dimensional surface known as the separatrix. It is sustained by a current sheet separating the dead magnetosphere with closed field lines from the active zone supported by open field lines. In an ideal plasma picture, this current sheet is infinitely thin, representing a discontinuity in the solution. In reality, due to some microphysics dynamics within the plasma, we expect this current sheet to spread around this separatrix with a depth depending on the plasma condition. Therefore, in a first stage, we assumed an ideally infinitely thin gap and in a second stage we release this assumption by using a thick slot gap.

\subsubsection{Sky maps and light curves}

\begin{figure}
	\centering
	\includegraphics[width=\columnwidth]{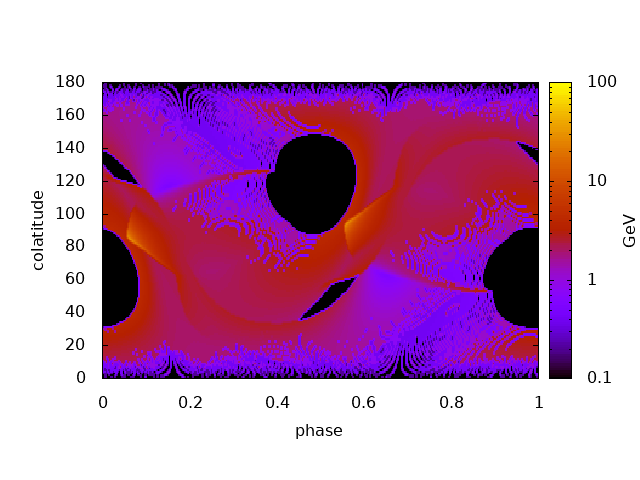} 
	\caption{\label{fig:carte_frequence_mink}Sky map showing the photon peak energies, in logarithmic scale, depending on the phase and line of sight for an obliquity $\rchi=60\degree$.}
\end{figure}

Let us start with the sky map in Fig.~\ref{fig:carte_frequence_mink} showing an example of the photon energy distribution received by a distant observer from the slot gaps, in logarithmic scale, the energy $E_{\rm c} = \hbar \, \omega_{\rm c}$ being obtained by multiplying the frequency of the observed radiation~$\omega_{\rm c}$ (including Doppler and aberration effects) by Planck reduced constant~$\hbar$. The parameters used assume an inclination angle of $\rchi=60\degree$ and a Lorentz factor $\gamma=10^{7}$ for the particles accelerated in these cavities, which are typical values quoted for example in \cite{becker_neutron_2009} and in \cite{beskin_physics_1993}. For each point on the map marked by the phase and inclination of the line of sight $(\phi,\zeta)$, in fact an area of size $0.5\degree\times0.5\degree$, only the energy of the most energetic photons at that point are displayed. This map therefore reveals the GeV photon production efficiency for each couple of phase-inclination $(\phi,\zeta)$ parameters, but without giving any indication of the actual shape of the spectra for each phase and each angle of the line of sight. We will detail these characteristics later.

Knowing the power of curvature radiation~$P_{\rm curv}$ in \eqref{eq:power} we calculate the number of photons emitted by each emission point along the magnetic field lines. In gamma-rays, because of the large span in energy range, we decided to cut the energy bands into evenly spaced intervals on a logarithmic scale. Each decade is cut into two intervals so that the successive values follow a geometric progression of reason $r=10^{1/2}\approx3.16$. The intervals are thus of the form $[r^{n},r^{n+1}]\,E_0$ with $n$ a positive integer and $E_0$ a characteristic energy set by default to~$E_0=1$~GeV. The following intervals are therefore chosen to represent the emission maps thus made for different obliquities~$\rchi$, as well as some associated light curves for different line of sight inclinations~$\zeta$: from 1 to $\sqrt{10}\approx3.16$ GeV as in Fig.~\ref{fig:carte_HE_1-3,16GeV}, from $\sqrt{10}\approx3.16$ to $10$~GeV as in \ref{fig:carte_HE_3,16-10GeV}, from $10$ to $\sqrt{100}\approx31.6$~GeV as in \ref{fig:carte_HE_10-31,6GeV} and from $\sqrt{100}\approx31.6$ to $100$~GeV as \ref{fig:carte_HE_31,6-100GeV}. Each of these maps takes into account only photons within a certain energy range, which are chosen as a geometric sequence as explained before. Most of the emission is concentrated between 3.16 and 10~GeV with a few high-energy emission points that can rise above 31.6~GeV. This feature will be detailed in the next paragraph about the spectra, demonstrating that most of the emitted photons are within the frequency range $[3,10]$~GeV with only a restricted area in the sky map that emits photons with an energy greater than $30$~GeV. 

\begin{figure}
	\centering
	\includegraphics[width=\columnwidth]{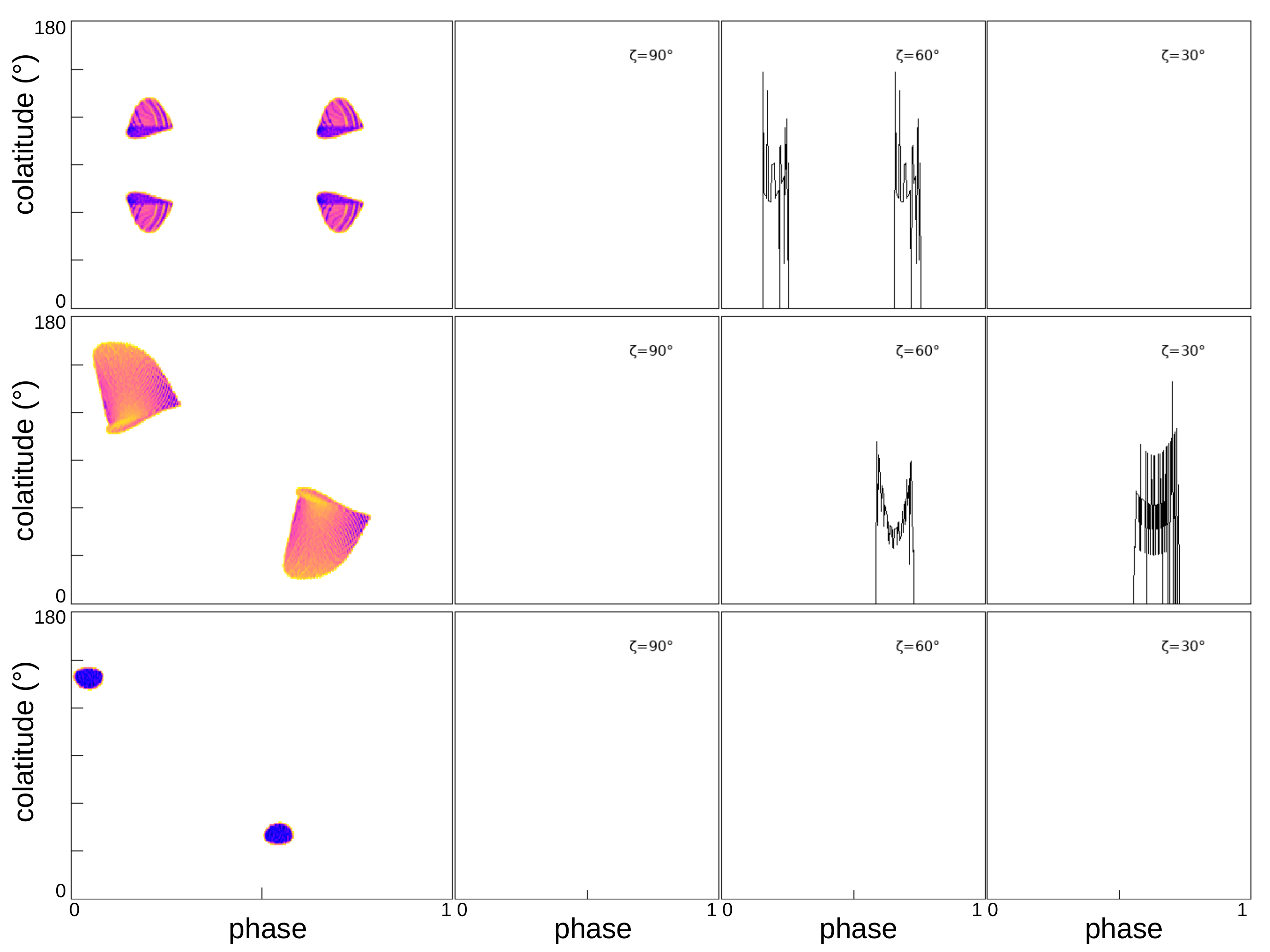}
	\caption{\label{fig:carte_HE_1-3,16GeV}Sky maps with photon energies in $[1,3.16]$~GeV for obliquities $\rchi=\{90\degree, 60\degree, 30\degree\}$ from top to bottom. Some light-curves are shown for several inclination angles $\zeta=\{90\degree, 60\degree, 30\degree\}$.}
\end{figure}
\begin{figure}
	\centering
	\includegraphics[width=\columnwidth]{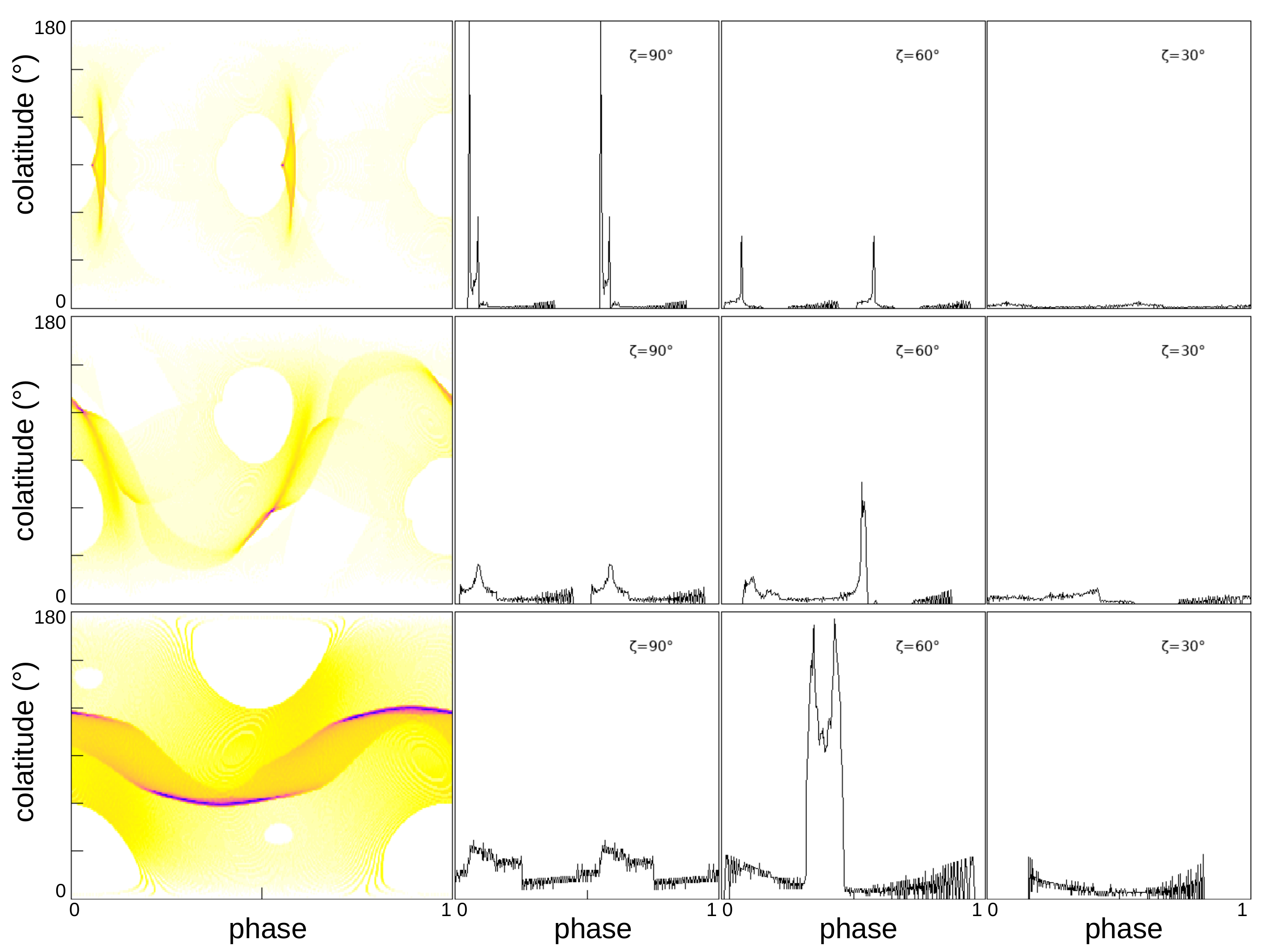}
	\caption{\label{fig:carte_HE_3,16-10GeV}Same as Fig.~\ref{fig:carte_HE_1-3,16GeV} but for photon energies between $[3.16,10]$~GeV.}
\end{figure}
\begin{figure}
	\centering
	\includegraphics[width=\columnwidth]{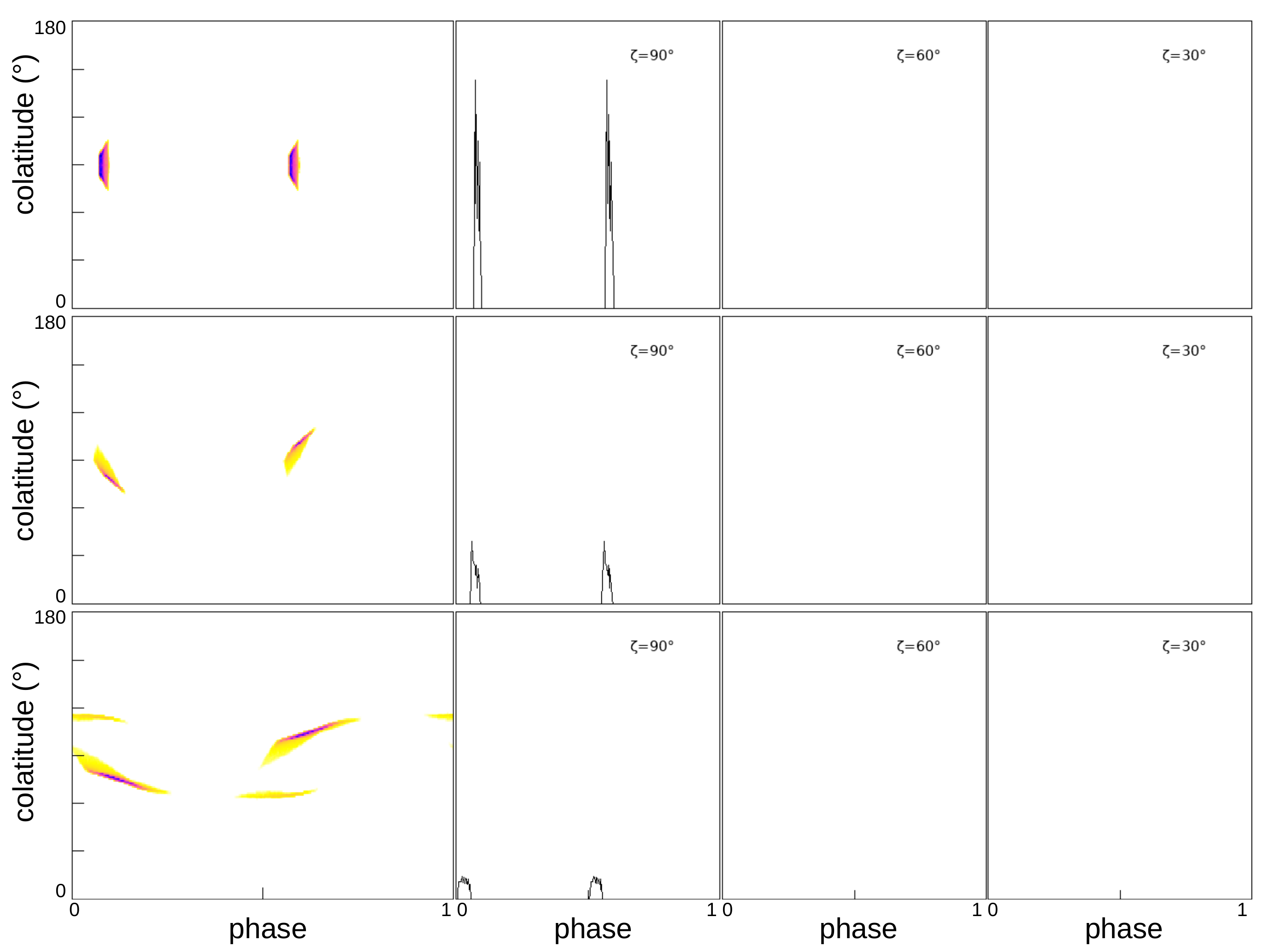}
	\caption{\label{fig:carte_HE_10-31,6GeV}Same as Fig.~\ref{fig:carte_HE_1-3,16GeV} but for photon energies between $[10,31.6]$~GeV.}
\end{figure}
\begin{figure}
	\centering
	\includegraphics[width=\columnwidth]{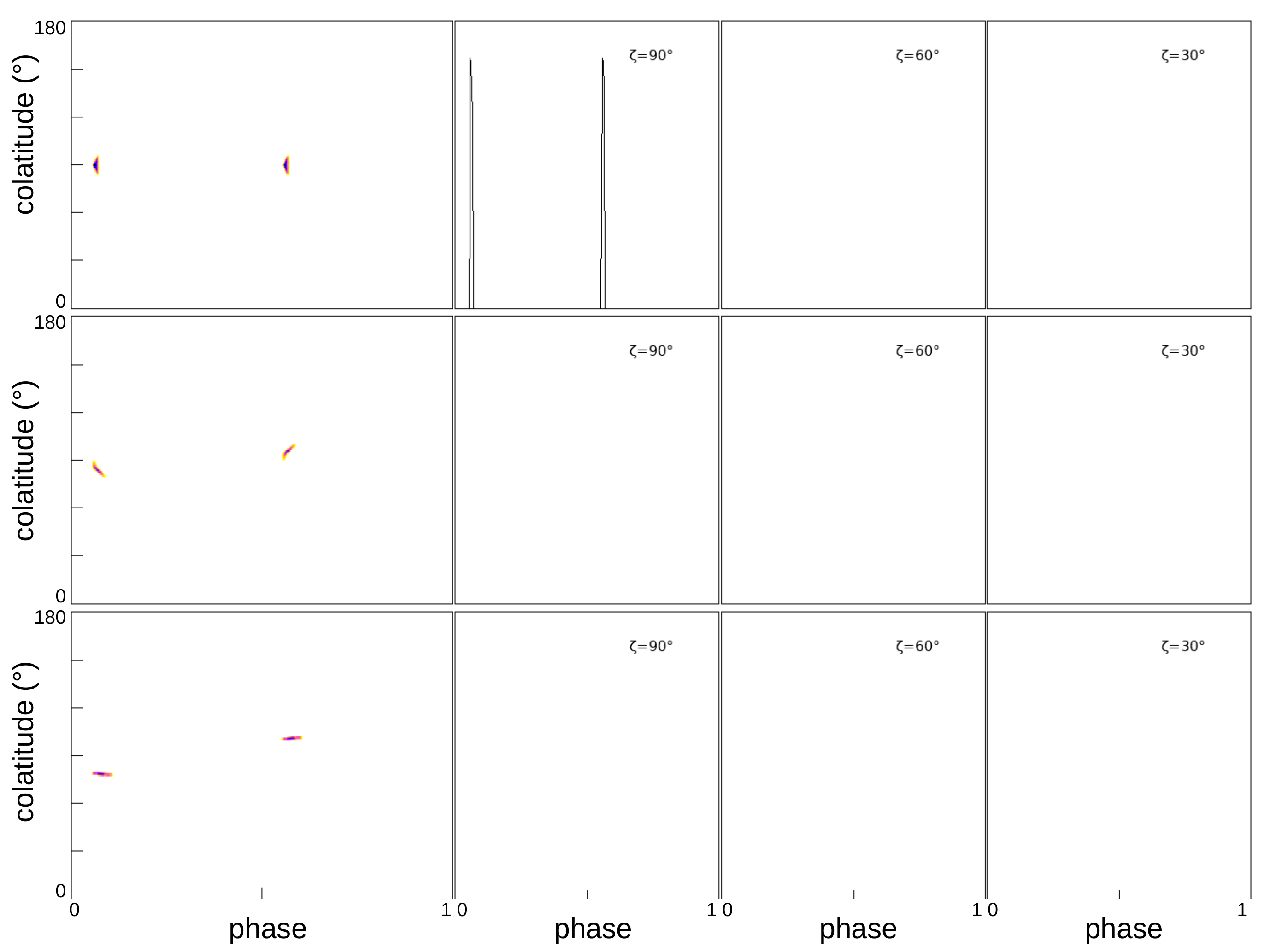}
	\caption{\label{fig:carte_HE_31,6-100GeV}Same as Fig.~\ref{fig:carte_HE_1-3,16GeV} but for photon energies between $[31.6,100]$~GeV.}
\end{figure}

\begin{figure}
	%\makebox[\columnwidth][c]{
	\includegraphics[width=\columnwidth]{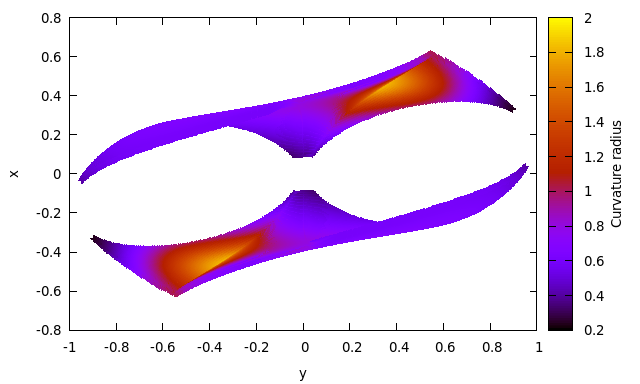}
	\caption{\label{img:carte_courb}Curvature radius depending on the position in the slot gaps for $\rchi=90\degree$ (projection on the xOy plane). The unit of length is the light-cylinder radius~$R_{\rm cyl}$).}
\end{figure}

\begin{figure}
	\includegraphics[width=\columnwidth]{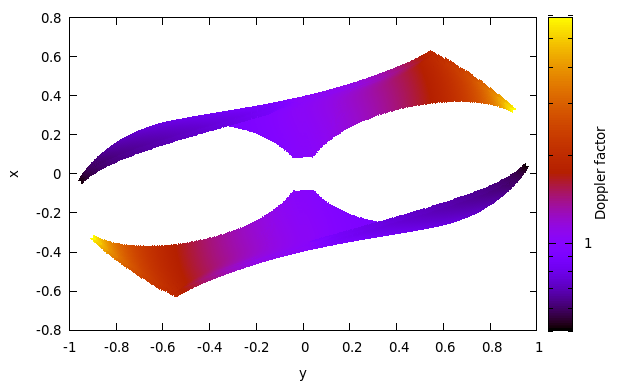}
	\caption{\label{img:carte_doppler} Doppler factor depending on the position in the slot gaps (projection on the xOY plane) for $\rchi=90\degree$.}
\end{figure}

\begin{figure}
	%\makebox[\columnwidth][c]{
	%\includegraphics[width=\columnwidth]{image/frequence90_log}%}
	\includegraphics[width=\columnwidth]{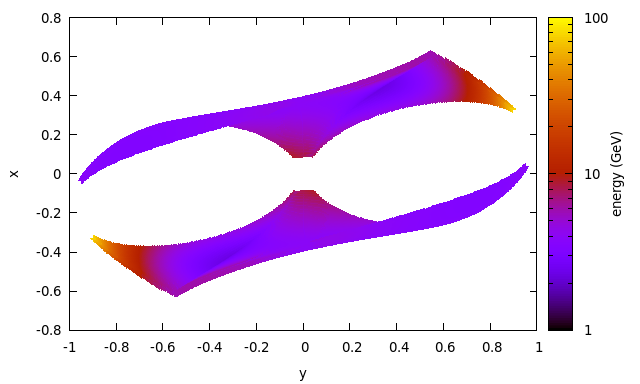}
	\caption{\label{img:carte_energ}Photon energy depending on the position in the slot gaps (projection on the xOy plane) for $\rchi=90\degree$.}
\end{figure}

Looking at these maps, only two regions stand out with very high energy photons, around 50~GeV, near the points $(\phi=0.1,\zeta=90\degree)$ and $(\phi=0.6,\zeta=90\degree)$. These areas are at high altitude, close to the light cylinder. The rest of the emission seems to show comparatively little variation in energy with the phase or inclination of the line of sight. The results of these maps are a non-trivial combination of the field line curvature which varies non-monotonically with altitude, see Fig~\ref{img:carte_courb}, and the Doppler factor producing a blue or red shift according to the line of sight geometry, see Fig~\ref{img:carte_doppler}. As a result, the most energetic photons are repelled towards the light cylinder, but only in regions for which the Doppler factor produces a blue shift, i.e. for particles moving along the field lines in the same direction as the direction of rotation, see Fig.~\ref{img:carte_energ}, particles moving in the opposite producing red shifted photons of lower energy.

\subsubsection{Spectra}

Looking at the spectra offers another perspective of pulsar multi-wavelength emission properties. So let us begin by studying the spectra of a thin slot gap without taking into account the weight assigned to each photon as a function of the radiated power \eqref{eq:power}.

\begin{figure}
	\centering
	\includegraphics[width=\columnwidth]{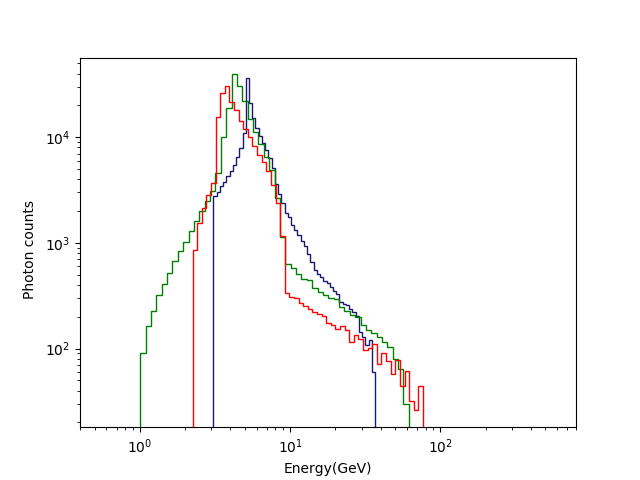} 
	\caption{\label{fig:Spectre_HE_sans_poids}Phase and line of sight averaged spectra for obliquity $\rchi=\{30\degree, 60\degree, 90\degree\}$ respectively in blue, green and red solid lines without curvature radiation power weights.}
	\vspace{1cm}
\end{figure}

The Fig.~\ref{fig:Spectre_HE_sans_poids} represents the phase and line of sight averaged spectra from these gaps for different obliquities~$\rchi=\{30\degree, 60\degree, 90\degree\}$ respectively in blue, green and red solid lines. We find a distribution consistent with what we saw in Fig.~\ref{fig:carte_frequence_mink} for a obliquity $\rchi=60\degree$ (green curve) with very few photons on the highest energies side, above 50~GeV and a high concentration of photons in the centre of the spectrum, around a few GeV. The range of the spectrum also varies significantly with the inclination of the magnetic dipole.

\begin{figure}
	\centering
	\includegraphics[width=\columnwidth]{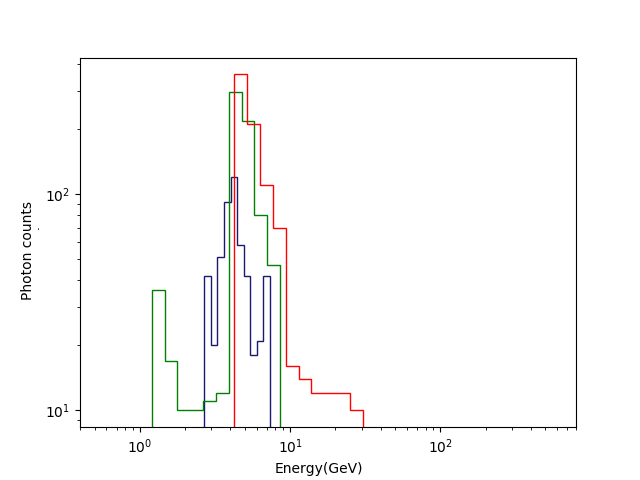} 
	\caption{\label{fig:Spectre_observateur_HE_sans_poids}Phase-integrated spectra for~$\rchi=60\degree$ and a line of sight inclination~$\zeta={30\degree,60\degree,90\degree}$ respectively in blue, green and red solid lines without curvature radiation weights.}
	\vspace{1cm}
\end{figure}

Nevertheless, from an observational as well as from a more realistic point of view, it is preferable not to average on the line of sight inclination angles since a particular observer only sees the part of the sky map in Fig.~\ref{fig:carte_frequence_mink} corresponding to a fixed angle~$\zeta$ (i.e. a horizontal line in this plot). An example of the variation of the phase-integrated spectrum (i.e. over the whole period of the pulsar) for $\rchi=60\degree$ is shown in Fig.~\ref{fig:Spectre_observateur_HE_sans_poids} for different values of the line of sight inclination $\zeta=\{30\degree,60\degree,90\degree\}$ again respectively in blue, green and red solid lines. We notice a great disparity in the shape and the limits of the spectra depending on the obliquity~$\rchi$ although the peak in the spectra remains at approximately the same energy, around 4~GeV. The smallest obliquity $\rchi=30\degree$ shows the tightest spectrum with an energy interval restricted to the band~$[3,7]$~GeV. A higher obliquity with $\rchi=60\degree$ moves this lower limit to about 1~GeV while the upper limit increases to 8-9~GeV. In addition, a double-peak spectrum appears, with one peak at low energy around 1.5~GeV and the other remaining at 4~GeV. We show later that taking into account the curvature radiation power~\eqref{eq:power} eliminates this second low energy peak. An even higher obliquity of $\rchi=90\degree$ shifts the spectrum in the opposite direction, towards the higher energies with a lower limit of 4~GeV and an upper limit of 30~GeV. As a conclusion, the observer line of sight strongly impacts the received phase-averaged spectrum. The behaviour of these spectra is a direct consequence of the geometry of the field lines, their visibility by the observer, and the aberration and Doppler effects.

The sky map in Fig.~\ref{fig:carte_frequence_mink} and the spectra in Fig.~\ref{fig:Spectre_HE_sans_poids} and \ref{fig:Spectre_observateur_HE_sans_poids} have been made on the assumption that for each emission point along the last closed magnetic field line a single photon is emitted, i.e. by applying the same procedure as in \cite{giraud_radio_2020}, the successive points being separated by a certain length~$\Delta s$ along the field line and arbitrarily fixed by the numerical integration code. A more realistic view must take into account the effectiveness of this curvature radiation, i.e. its power spectrum. Indeed, since we know the radius of curvature of the magnetic field lines at each point, we can deduce the power radiated by a particle accelerated along them using the equation~\eqref{eq:power}. For the moment, we have not indicated the density of particles in the emitting regions, so we cannot calculate a precise flux detected on Earth, but we can give as an indication the distribution~$E\,dN/dE$ in energy of photons produced by the curvature radiation. The units remain arbitrary, we will write them down in~$UA$.

\begin{figure}
	\centering
	\includegraphics[width=\columnwidth]{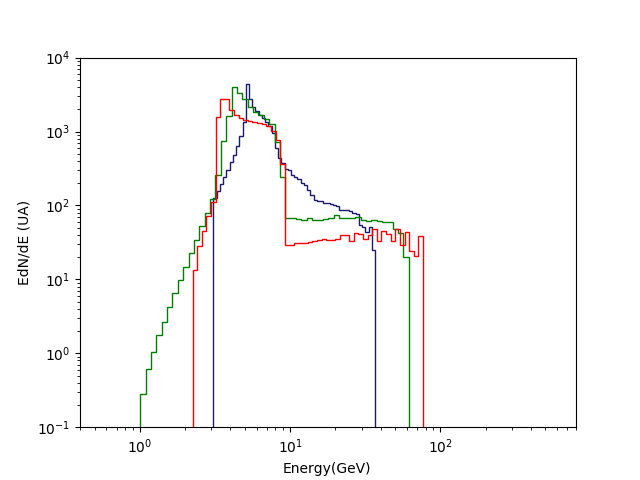}
	\caption{\label{fig:spectre_HE}Same as Fig.\ref{fig:Spectre_HE_sans_poids} but with curvature radiation power included.} 
\end{figure}

Taking the same approach as above, examples of spectra are shown in Fig.~\ref{fig:spectre_HE}. As only one value of the Lorentz factor is used, these spectra show rather abrupt decreases and increases in intensity as a function of frequency. As in the spectrum of Fig.~\ref{fig:Spectre_HE_sans_poids}, a weak flux of photons is produced with an energy above $50$~GeV and a strong flux of photons for an energy in the vicinity of $5$~GeV. Taking into account the modulation of the intensity of the spectrum as a function of the locally radiated power does not change the frequency range of the emitted photons. This is why the lower and upper limits of the spectra are identical in Fig.~\ref{fig:Spectre_HE_sans_poids} and Fig.~\ref{fig:spectre_HE}. Indeed, the radiated power changes the number of photons produced but not their energy.

\begin{figure}
	\centering
	\includegraphics[width=\columnwidth]{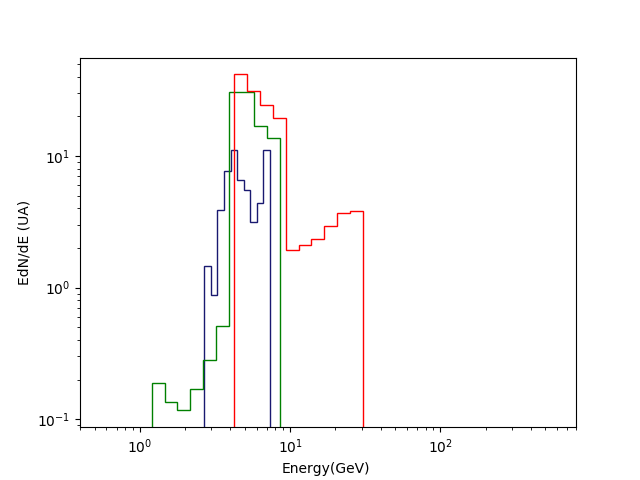} 
	\caption{\label{fig:Spectre_observateur_HE_avec_poids}Same as Fig.\ref{fig:Spectre_observateur_HE_sans_poids} but with curvature radiation power included.}
%	\vspace{1cm}
\end{figure}

As in  Fig.\ref{fig:Spectre_HE_sans_poids}, these spectra are averaged over the phase and line of sight inclination. A real observer will only measure the spectrum associated with a fixed value of the line of sight~$\zeta$. Following the same procedure as for Fig.\ref{fig:spectre_HE}, another example of a phase-integrated spectrum taking into account the power of curvature radiation is shown in Fig~\ref{fig:Spectre_observateur_HE_avec_poids} for $\rchi=60\degree$ and for different lines of sight $\zeta=\{30\degree,60\degree,90\degree\}$, respectively in blue, green and red solid lines. The limits of the spectra remain identical to those of Fig.~\ref{fig:Spectre_observateur_HE_sans_poids} since the energy of the radiated photons does not change, only their number is modified by taking into account the curvature radiation power. The spectrum in blue for $\zeta=30\degree$ is very similar to its counterpart without taking into account the power. On the other hand, for $\zeta=60\degree$ in green, the low-energy peak has disappeared to show only an intense peak around 4-8~GeV, in the form of a plateau. Finally, for $\zeta=90\degree$, the high-energy component around 30~GeV stands out very clearly and becomes comparable to the component around 4-8~GeV. This example underlines the importance of the radiated power on the weighting of the real spectrum of a pulsar. It is not enough to simply count the photons emitted in isolation at each point.

\begin{figure*}
\centering
\begin{tabular}{cc}
	\includegraphics[width=0.5\columnwidth]{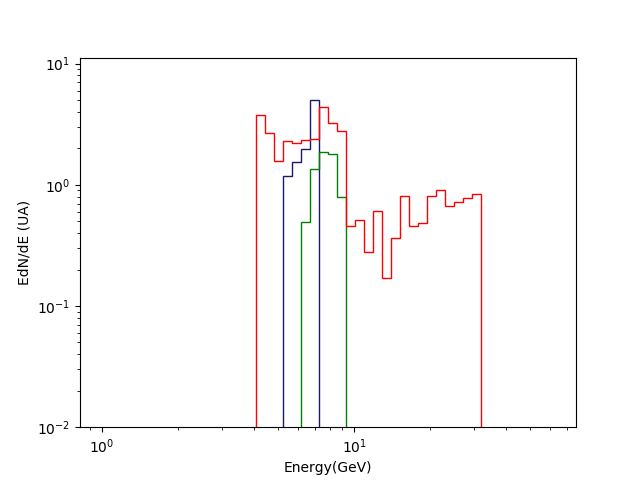} &
	\includegraphics[width=0.5\columnwidth]{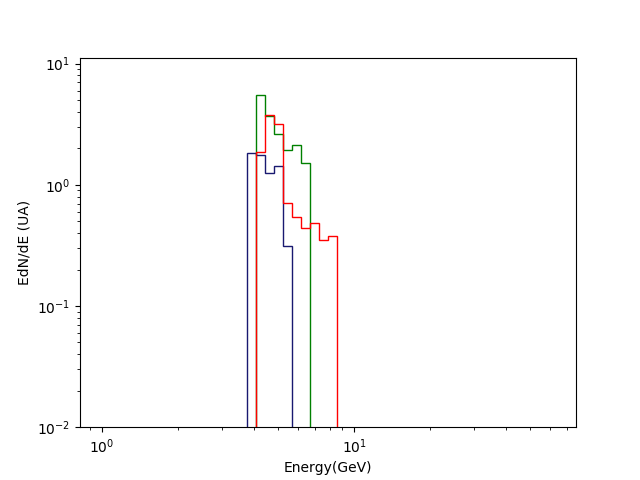}
	\vspace{-4mm} \\
	\includegraphics[width=0.5\columnwidth]{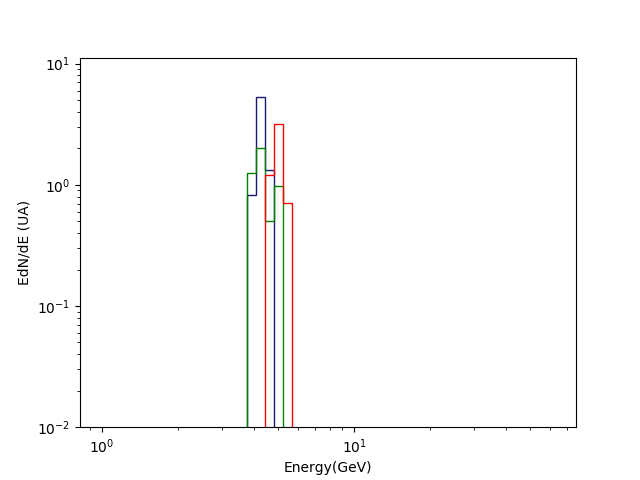} &
	\includegraphics[width=0.5\columnwidth]{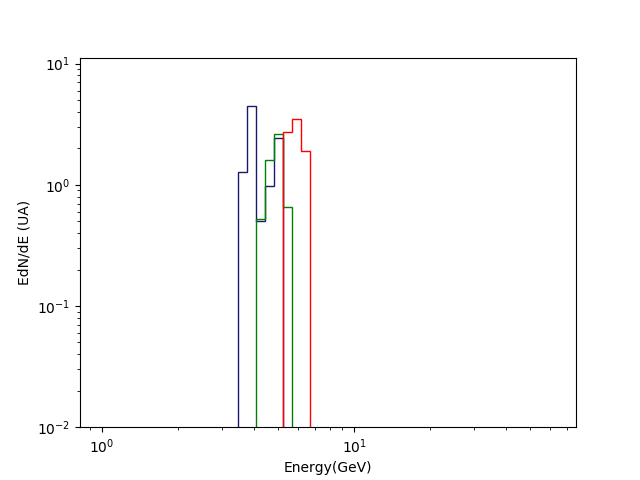}
	\vspace{-4mm} \\
	\includegraphics[width=0.5\columnwidth]{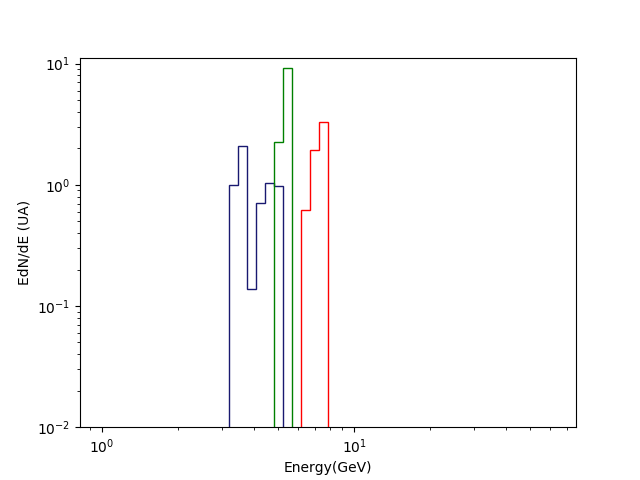} &
	\includegraphics[width=0.5\columnwidth]{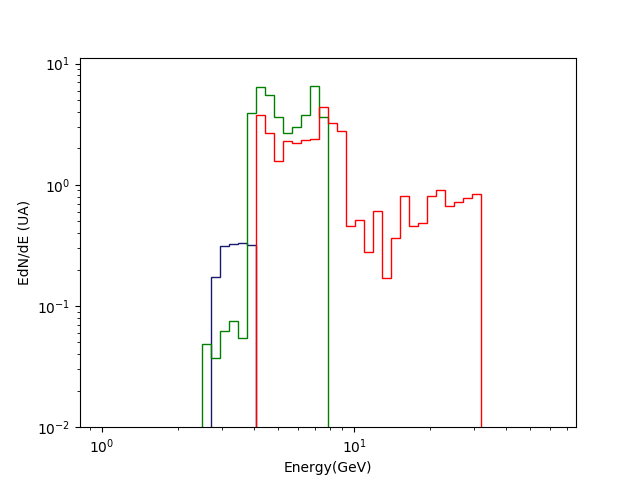}
	\vspace{-4mm} \\
	\includegraphics[width=0.5\columnwidth]{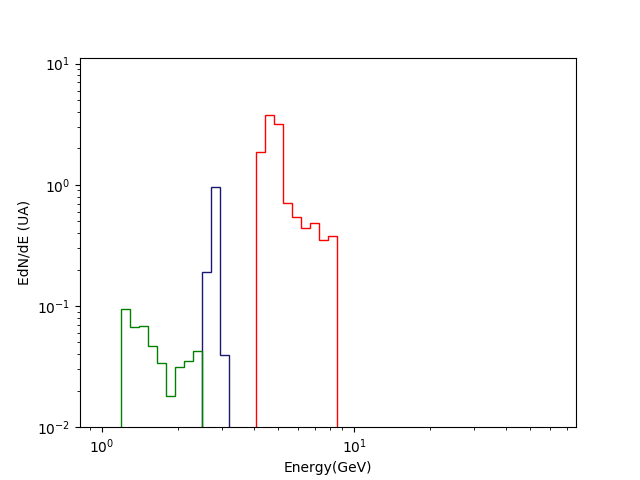} &
	\includegraphics[width=0.5\columnwidth]{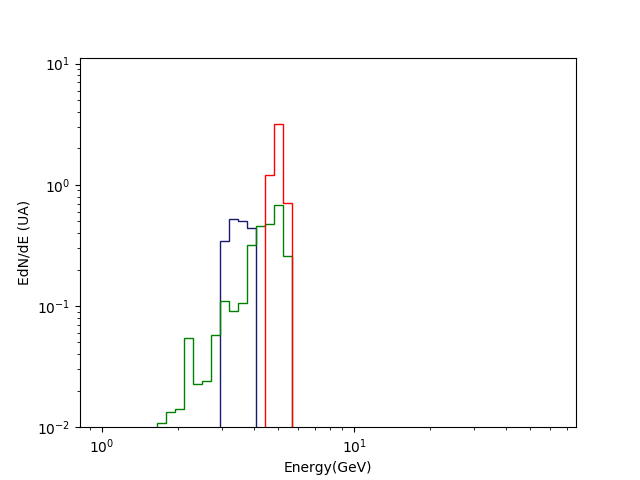}
	\vspace{-4mm} \\
	\includegraphics[width=0.5\columnwidth]{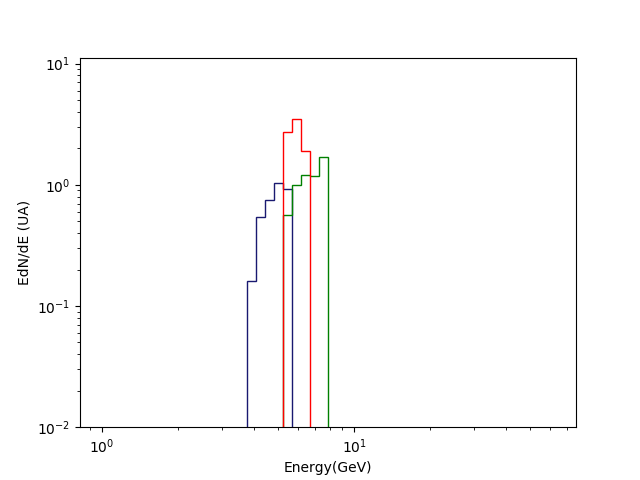} &
	\includegraphics[width=0.5\columnwidth]{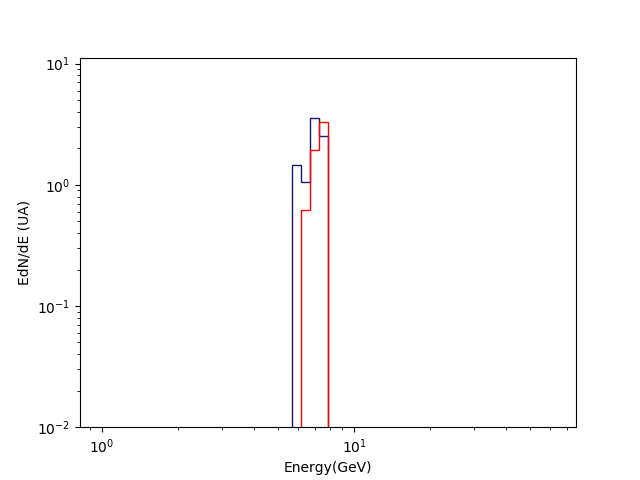}
	\end{tabular}
	\caption{\label{fig:Spectre_60_observateur_1-10} Phase resolved spectra for the phase $\phi$ in regular intervals of length 0.1 and subdivided into $[k,0.1\,(k+1)]$ with $k\in[0..9]$ with~$\rchi=60\degree$ and $\zeta=\{30\degree,60\degree,90\degree\}$ respectively in blue, green and red for a thin slot gap.}
\end{figure*}

Applying these methods for curvature radiation, we are able to compute phase-resolved spectra for a given magnetic axis inclination~$\rchi$ and line of sight~$\zeta$ depicting a particular observer. Performing this calculation, the phase-integrated spectra have been divided into 10~regular phase intervals~$\Delta \phi$ each of length equal to 10\% of the period thus $\Delta t = 0.1\,P$ or $\Delta \phi = 36\degree$. Figs.~\ref{fig:Spectre_60_observateur_1-10} shows an example of phase-resolved spectra according to this breakdown of the period. 
The phases correspond to the regular intervals from $[0.0.1]$ to $[0.9,1.0]$. For the first phase interval, we already see a strong variation in the shape of the spectrum for a line of sight inclined at $\zeta=90\degree$, represented by the red curves. Photon energy rises to over 30~GeV in the range~$[0.0.1]$ and falls below 9~GeV in the next range~$[0.1.0.2]$. These variations are much less noticeable at lower inclinations of $\rchi = 30\degree$, the blue curves on our spectra, and $\rchi=60\degree$, the green curves. In the phase interval $[0.2,0.3]$  all the spectra shrink significantly with a narrow energy peak around 5~GeV regardless of the inclination~$\zeta$ on the left curve. Then the spectrum widens again slightly towards the higher energies on the interval $[0.3,0.4]$. The separation of these spectra becomes more and more visible in the following phases, for example in $[0.4,0.5]$. Energies above 30~GeV dominate again for $\zeta=90\degree$ in the $[0.5,0.6]$ phase while the other two spectra widen towards the low energies. The separation of the spectra becomes very clear again in phase $[0.6,0.7]$. At the next phase in $[0.7,0.8]$, the spectra converge more and more to refocus around 7~GeV.
This trend continues over the next phase intervals, $[0.8,0.9]$ and $[0.9,1]$ with a disappearance of the pulsation for $\zeta=60\degree$ in phase $[0.9,1.0]$ because it corresponds to the location of the polar cap shadow, visible in Fig.~\ref{fig:carte_frequence_mink}, emission disappearing.

Note that the maximum intensity of the spectrum is not very sensitive to the phase interval, whatever the inclination and phase considered, the maximum of the spectra is between 1 and 10 in our arbitrary units. Only the inclination of $\zeta=60\degree$ seems to show a strong amplitude or even a disappearance of the signal for some given phase intervals.

\subsection{Thick slot gap}

Thin slot gaps are idealized regions of infinitely small thickness producing very sharp gamma-ray peaks due to caustics effects \citep{morini_inverse_1983}. Until now we only considered emission along the last closed field lines, neglecting a possible thickness of the current sheet. However, because of the plasma and radiation dynamics within these gaps, we expect an emission layer of finite thickness to form. Therefore, in this paragraph we consider slot gaps relying not only on last closed field lines but also on field lines in their vicinity, some crossing the light cylinder, being open lines, and some staying within the light-cylinder, being closed lines.
Therefore, in order to tend to a more realistic description of the pulsar high-energy emission from slot gaps, we added in our simulations emission coming from magnetic field lines slightly above or slightly below the last closed field lines.

Sky maps and spectra shown in this paragraph have been realised assuming a thick layer of emission along the last closed magnetic field lines. The power radiated by particles is multiply by a Gaussian weight~$w_{\rm g}$ so the emission is maximum along the last closed field lines and decrease when there is an increase in distance between the line where the emission take place and the last closed field line. The colatitude of the last closed field line foot point on the stellar surface being marked by $\theta_{\rm pc}$ and the colatitude of any field line by $\theta$, in the coordinate system aligned with the magnetic field axis, the Gaussian weight is written as
\begin{equation}
w_{\rm g} = e^{-(\theta-\theta_{\rm pc})^{2}/\delta^{2}} .
\label{eq:gauss_HE}
\end{equation}
We take $\delta=\Delta\theta/5$ with an angular distance between two successive magnetic foot points at the same longitude~$\varphi$ equal to $\Delta\theta=\pi/100$.

\subsubsection{Sky maps and light curves}

\begin{figure}
	\centering
	\includegraphics[width=\columnwidth]{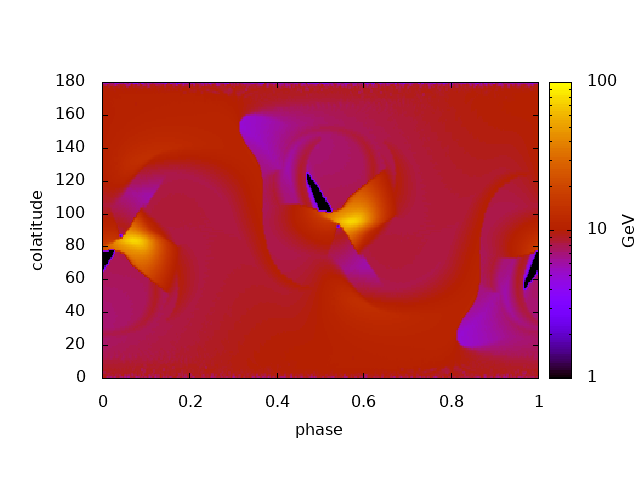} 
	\caption{\label{fig:carte_frequence_mink_epais}Sky map showing the photon peak energies, in logarithmic scale, depending on the phase and line of sight for an obliquity $\rchi=60\degree$.}
\end{figure}

The sky map of the highest photon energy is shown in fig.~\ref{fig:carte_frequence_mink_epais} for thick slot gaps and an obliquity~$\rchi=60\degree$. This figure must be compared with fig.~\ref{fig:carte_frequence_mink} for a thin gap. Now the most energetic photons are still located around the same phase and line of sight inclination with maximum energy around 100~GeV, but as expected the spreading of the current sheet extends the visibility region of high energy radiation in this sky map. The shadows of the polar caps have almost disappeared because their location overlaps with thick slot gap emission pattern.

\begin{figure}
	\includegraphics[width=\columnwidth]{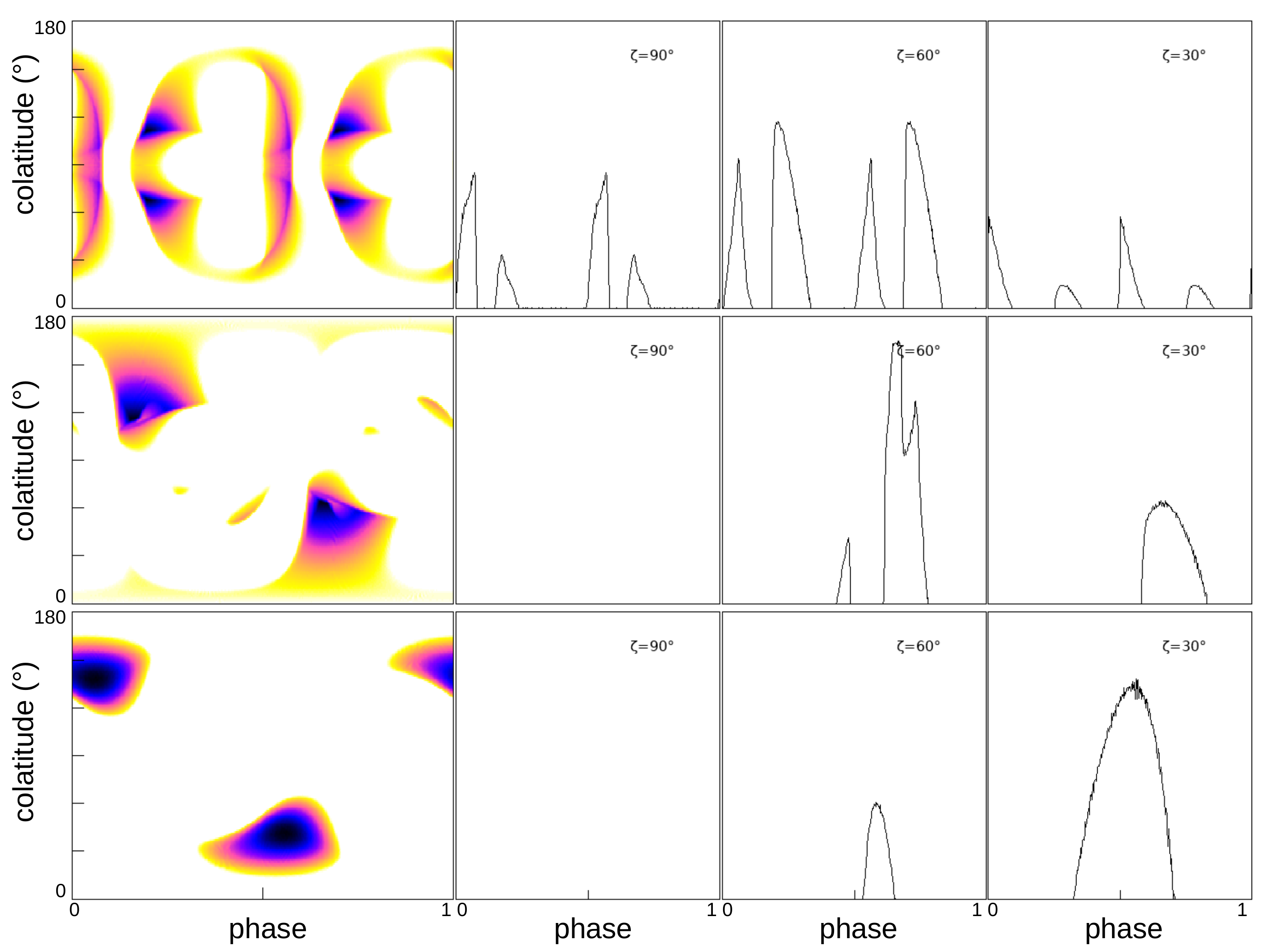} 
	\caption{\label{lct1-3GeV}Same as Figs.~\ref{fig:carte_HE_1-3,16GeV} but for thick slot gaps.} 
\end{figure}

\begin{figure}
	\includegraphics[width=\columnwidth]{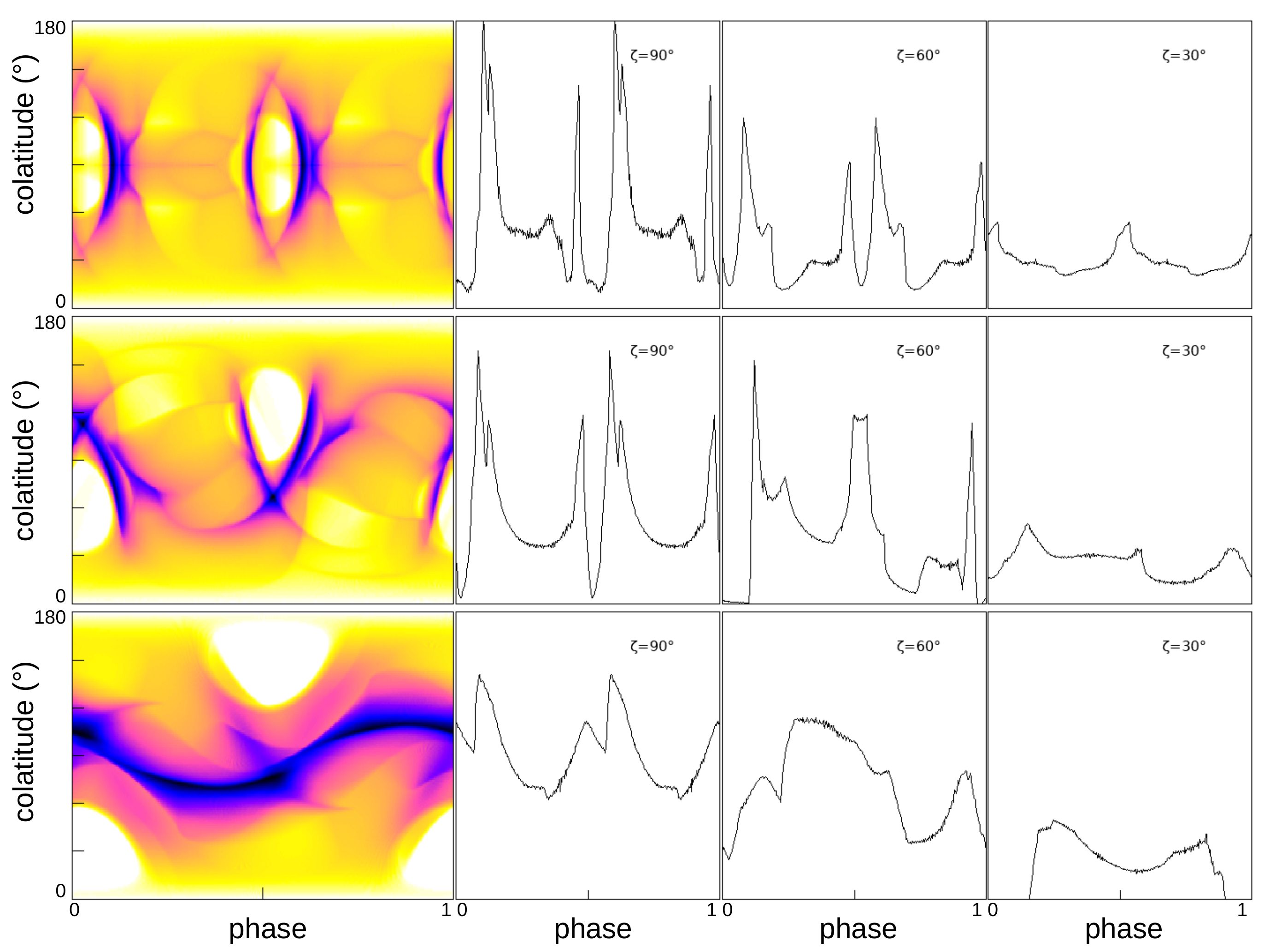} 
	\caption{\label{lct3-10GeV}Same as Figs.~\ref{fig:carte_HE_3,16-10GeV} but for thick slot gaps.} 
\end{figure}

\begin{figure}
	\includegraphics[width=\columnwidth]{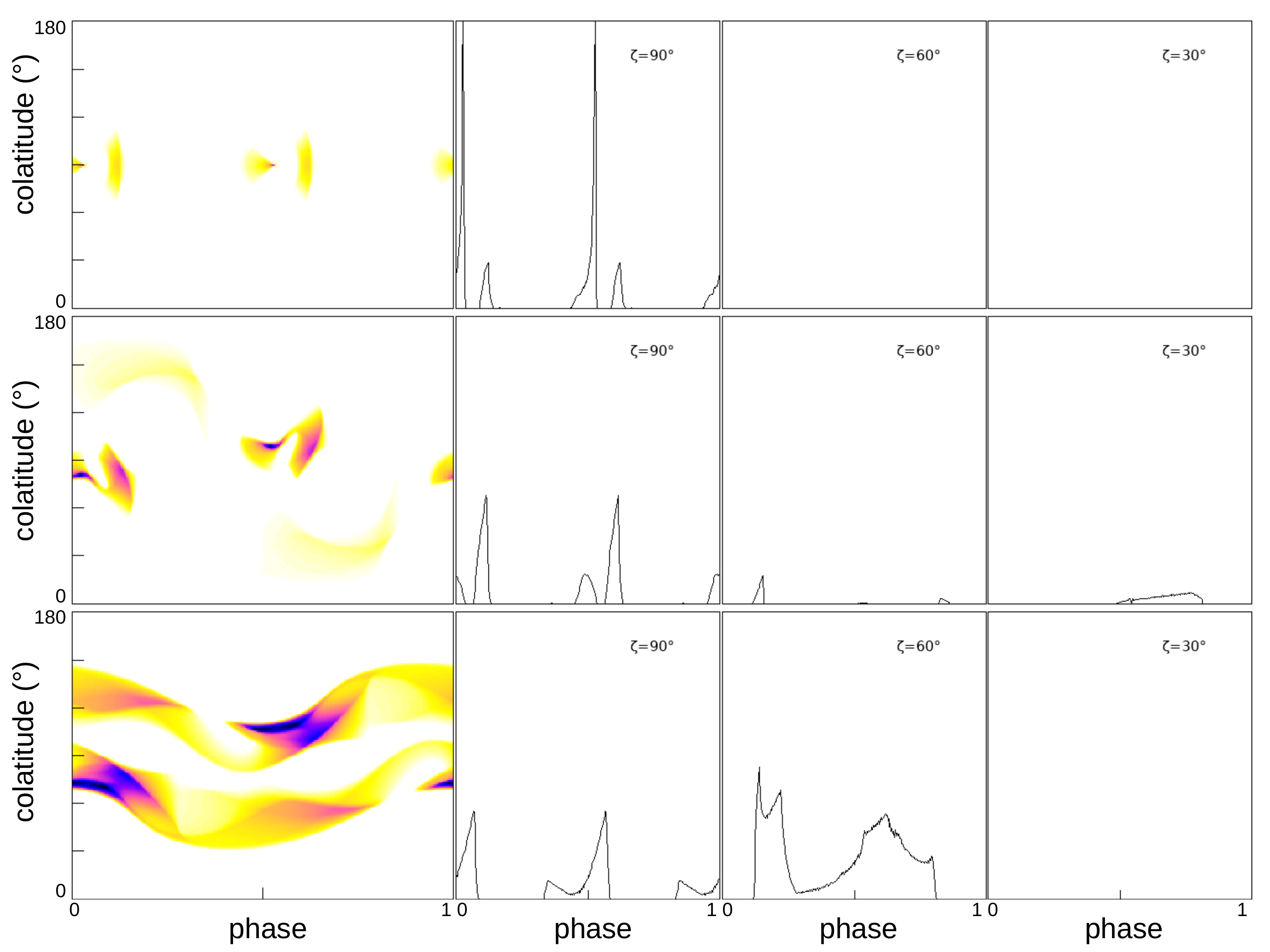} 
	\caption{\label{lct10-30GeV}Same as Figs.~\ref{fig:carte_HE_10-31,6GeV} but for thick slot gaps.} 
\end{figure}

\begin{figure}
	\includegraphics[width=\columnwidth]{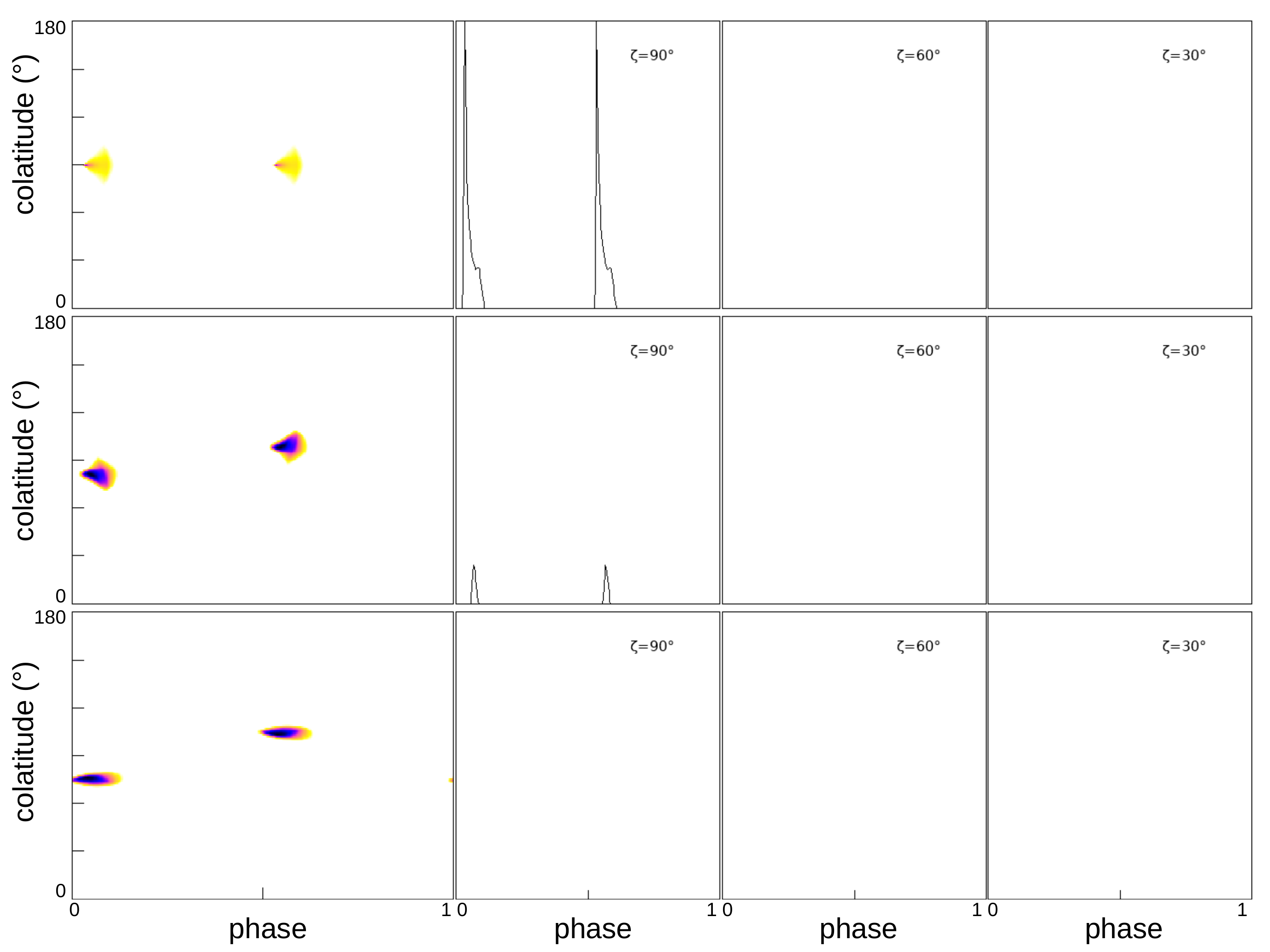} 
	\caption{\label{lct30-100GeV}Same as Figs.~\ref{fig:carte_HE_31,6-100GeV} but for thick slot gaps.} 
\end{figure}

Equivalently, figs.~\ref{lct1-3GeV} to \ref{lct30-100GeV} represent the sky maps for thick slot gaps for different energy ranges, the same as those shown in Figs.~\ref{fig:carte_HE_1-3,16GeV} to \ref{fig:carte_HE_31,6-100GeV}, and different $\rchi$ inclinations of the magnetic axis. Due to the finite thickness of the gap, the light-curves broaden in relation to the transverse size of the slot gaps. At any energies, the bright part of the sky maps has enlarged to a significant part of the full sky. Moreover, the light-curves show complex multi pulse profiles varying with energy. However, as already stated before, these sky maps and light curves samples serve as building blocks for the elaboration of more realistic and complete radiation patterns in pulsar magnetospheres. The particle distribution function and the full curvature spectrum must be taken into account. This is why we postponed a meaningful comparison with radio and gamma-ray observations for future works.

\subsubsection{Spectra}

\begin{figure}
	\centering
	\includegraphics[width=\columnwidth]{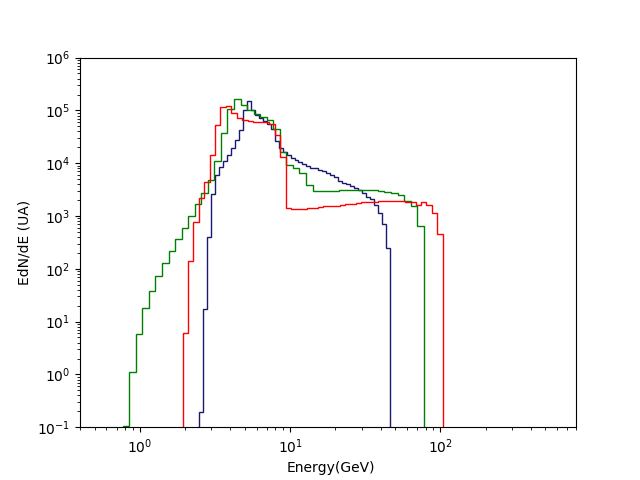} 
	\caption{\label{fig:spectre_HE_epais}Same as Fig.~\ref{fig:spectre_HE} but for a thick emission region.}
\end{figure}

The emission spectra and maps computed in the previous section have been made with an infinitely thin emission area along the last open field lines. Fig.~\ref{fig:spectre_HE_epais} shows the spectrum of the high energy emission for an emission area of a given thickness, using the same method as that used to draw the sky maps in Fig.~\ref{fig:spectre_HE}.
%\begin{figure}
%	\includegraphics[width=\columnwidth]{images/Spectre_HE_epais.png} 
%	\caption{\label{fig:spectre_flux_HE_epais}Same as Fig.~\ref{fig:spectre_HE} but for a thick emission region.}
%\end{figure}
According to this spectrum, when a thickness transverse to the emission area is included, additional radiation is received including lower and higher photon energies than those of a thin gap. To the spectrum of the infinitely thin zone are thus added other spectra from neighbouring field lines but whose curvature varies slightly, increasing or decreasing according to the nature of the field line: open or closed, which will obviously lead to variations in the radiated power and energy of the radiation. The differences in the shape of the spectra in Figs.~\ref{fig:spectre_HE} and \ref{fig:spectre_HE_epais} can also be partly explained by the modulation of intensity under the influence of the Gaussian factor which we have introduced by the weight~\eqref{eq:gauss_HE} and which has also been used here. Thus for each point of impact on the celestial sphere, the measured intensity is the power radiated by the particle multiplied by this Gaussian function.

\begin{figure*}
\centering
\begin{tabular}{cc}
	\includegraphics[width=0.5\columnwidth]{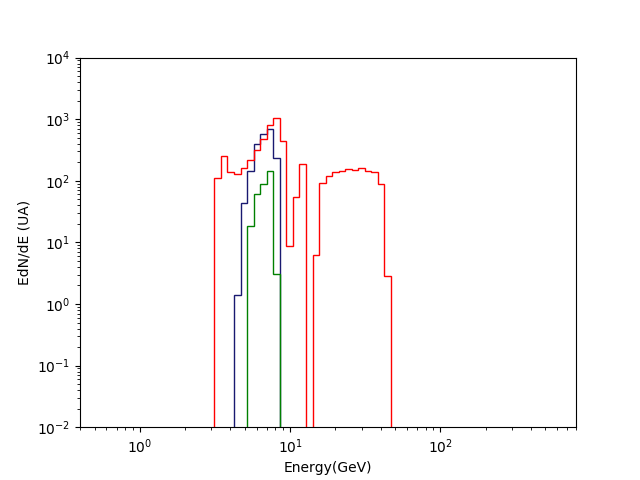} &
	\includegraphics[width=0.5\columnwidth]{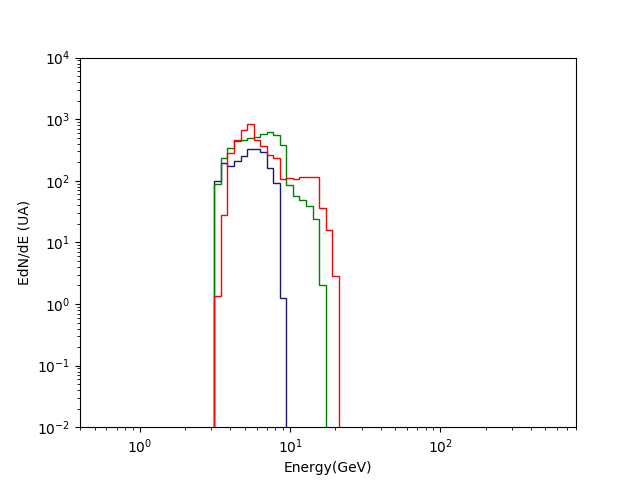}
	\vspace{-3mm}  \\
	\includegraphics[width=0.5\columnwidth]{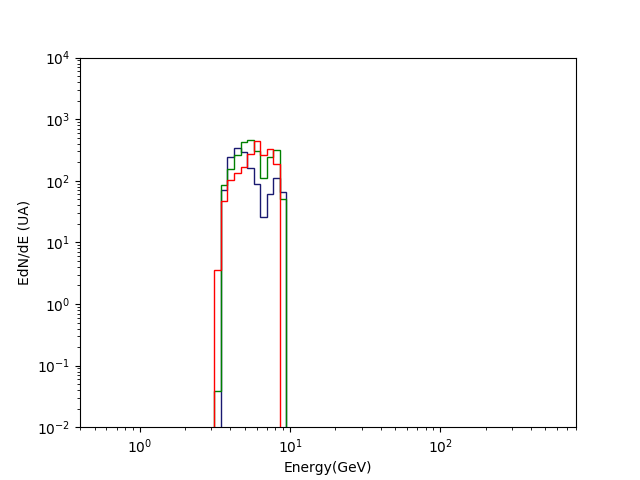} &
	\includegraphics[width=0.5\columnwidth]{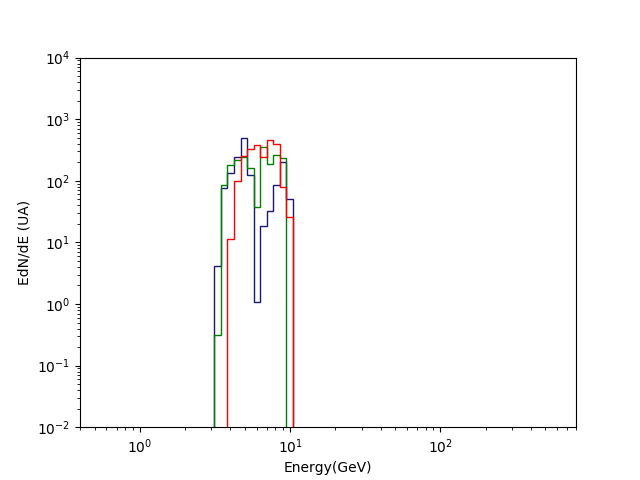} 
	\vspace{-3mm} \\
	\includegraphics[width=0.5\columnwidth]{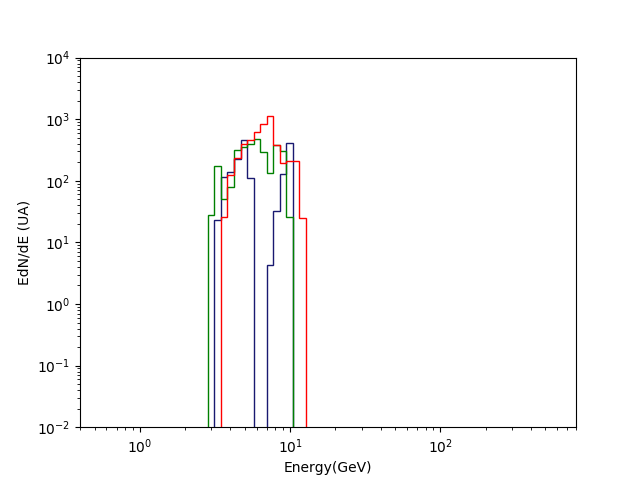} &
	\includegraphics[width=0.5\columnwidth]{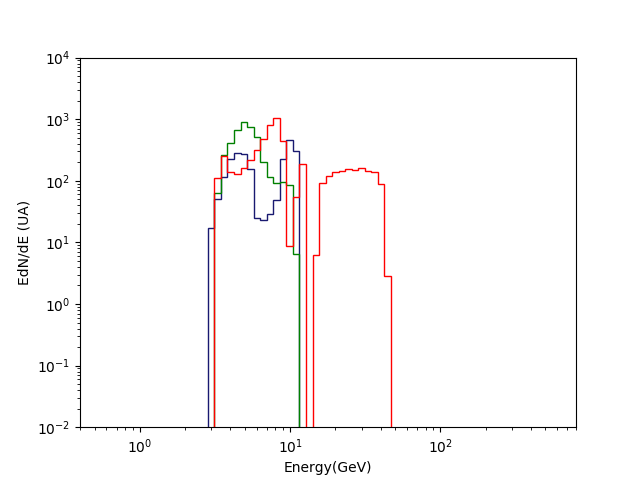} 
	\vspace{-3mm} \\
	\includegraphics[width=0.5\columnwidth]{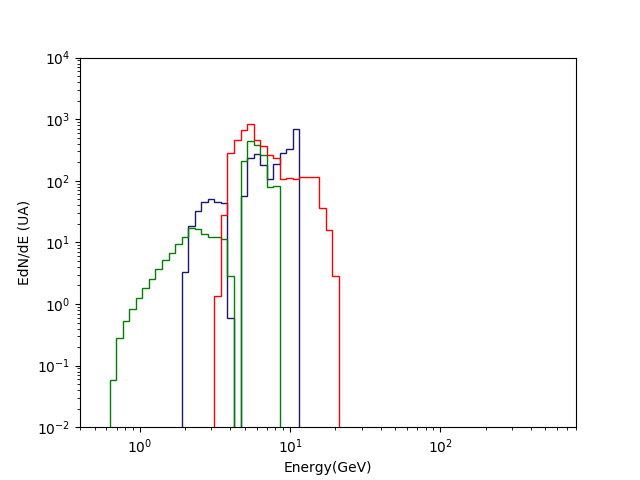} &
	\includegraphics[width=0.5\columnwidth]{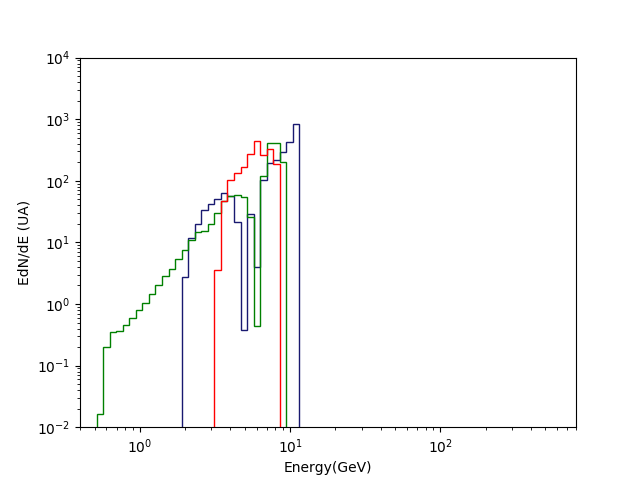} 
	\vspace{-3mm} \\
	\includegraphics[width=0.5\columnwidth]{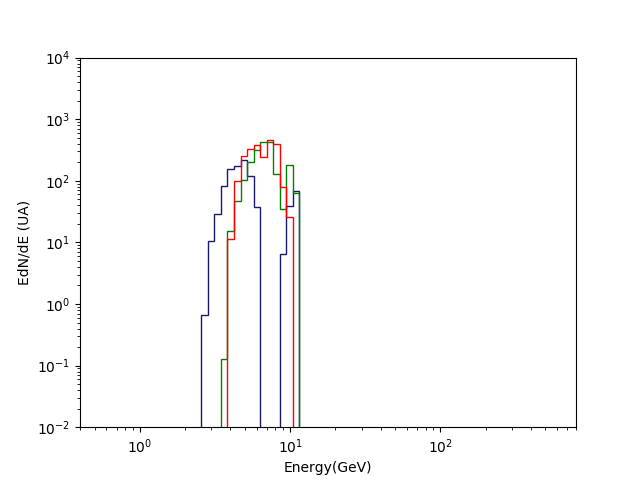} &
	\includegraphics[width=0.5\columnwidth]{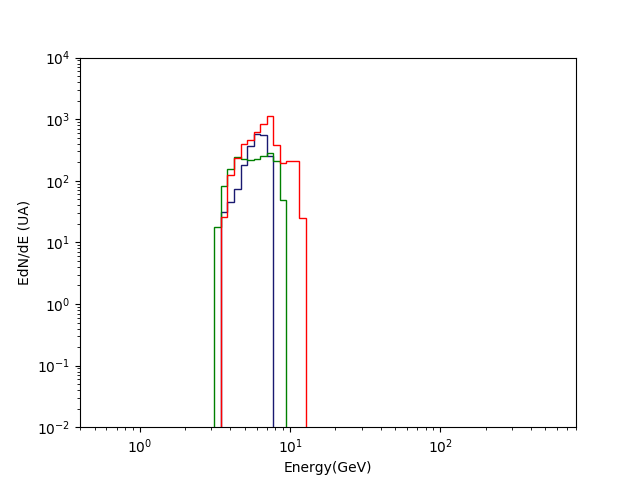}
	\end{tabular}
	\caption{\label{Spectre_60_observateur_1-10_epais}Same as Fig.~\ref{fig:Spectre_60_observateur_1-10} but for a thick slot gap.}
\end{figure*}

The Fig~\ref{Spectre_60_observateur_1-10_epais} represents the evolution of the high energy spectra over different phase intervals for an inclination $\chi=60\degree$ and several fixed values of the observation angle $\zeta$. It is the same as in Fig.~\ref{fig:Spectre_60_observateur_1-10} but for thick slot gaps. The emission is visible for all observation angles and for all phases unlike the spectra obtained for radiation without thickness of the emission zone, notably for the phase $[0.9,1]$, because, as seen in Fig.\ref{lct3-10GeV}, high energy emission is received at the phase of the polar cap shadow.

\begin{figure}
	\centering
	\includegraphics[width=\columnwidth]{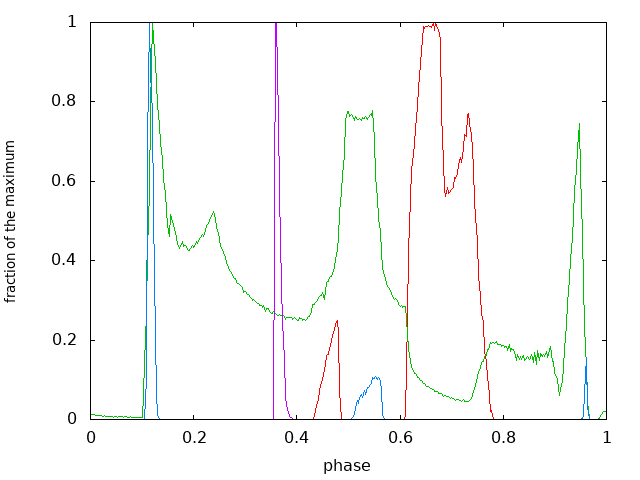} 
	\caption{\label{lc_HE_60-60_epais} Light curves for $\chi=\zeta=60\degree$ for different energy intervals: in red $[1, 3.16]$~GeV, in green $[3.16,10]$~GeV, in blue $[10,31.6]$~GeV and in violet $[31.6,100]$~GeV.}
\end{figure}

Fig.~\ref{lc_HE_60-60_epais} represents the evolution of the shape of the light curve observed for $\chi=\zeta=60\degree$ for different energies, the light curves being normalized by its maximum in each plot. It highlights the variety of pulse shape and width depending on the observed energy range.

We will now determine from which altitude in the magnetosphere of the pulsar this high-energy pulsed emission originates.

%\subsection{Mono-energetic particle spectrum}

\subsection{Altitude of emission}

\begin{figure}
	\includegraphics[width=\columnwidth]{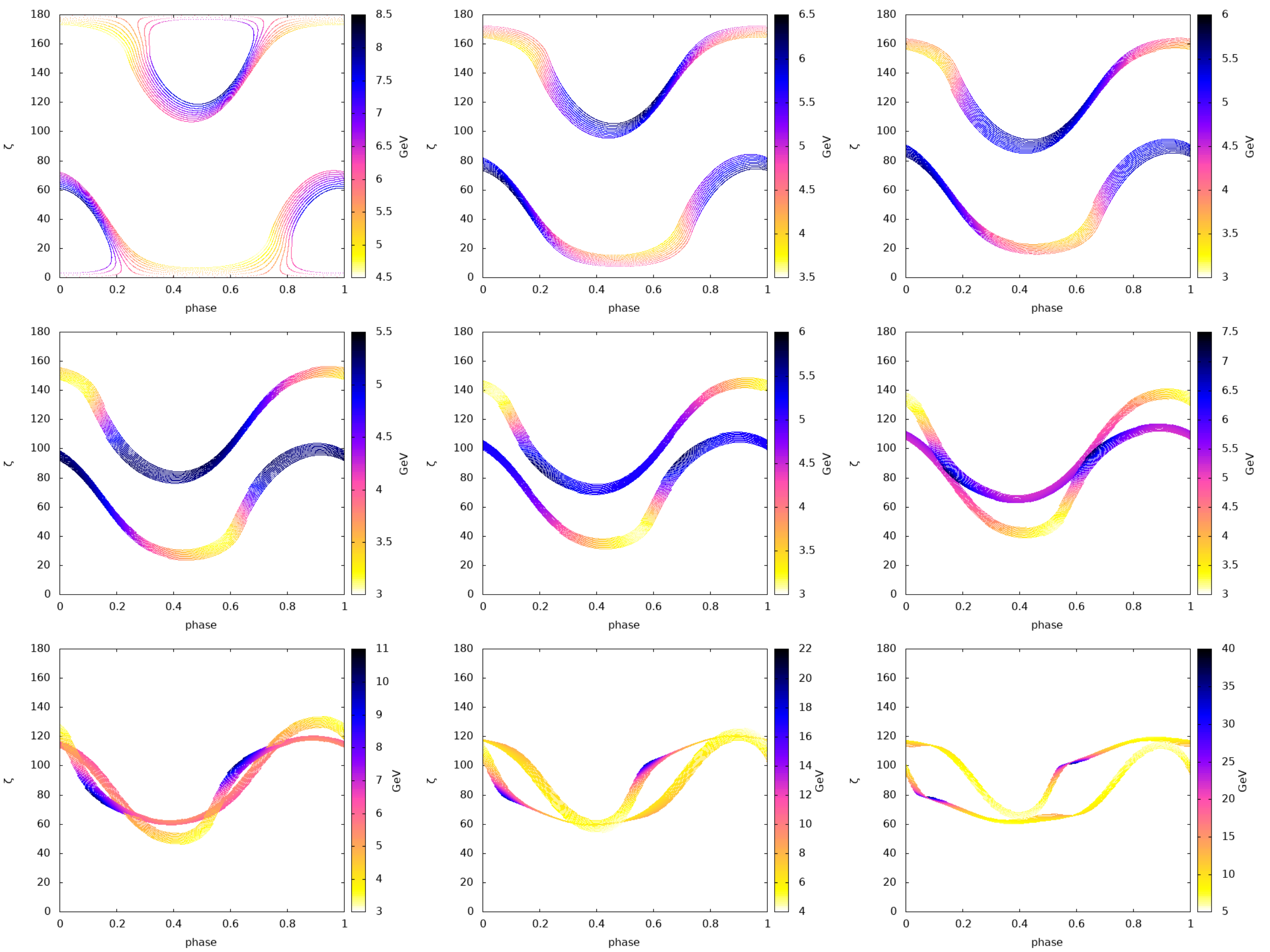}
	\caption{\label{alt30}Sky maps of photon energy for an inclination $\rchi=30\degree$, each map represents the contribution from a piece of the magnetosphere : those located in a spherical shell of thickness $R_{\star}$ whose altitude is increased by $R_{\star}$ to the next map.}
\end{figure}

\begin{figure}
	%	\makebox[\columnwidth][c]{
	\includegraphics[width=\columnwidth]{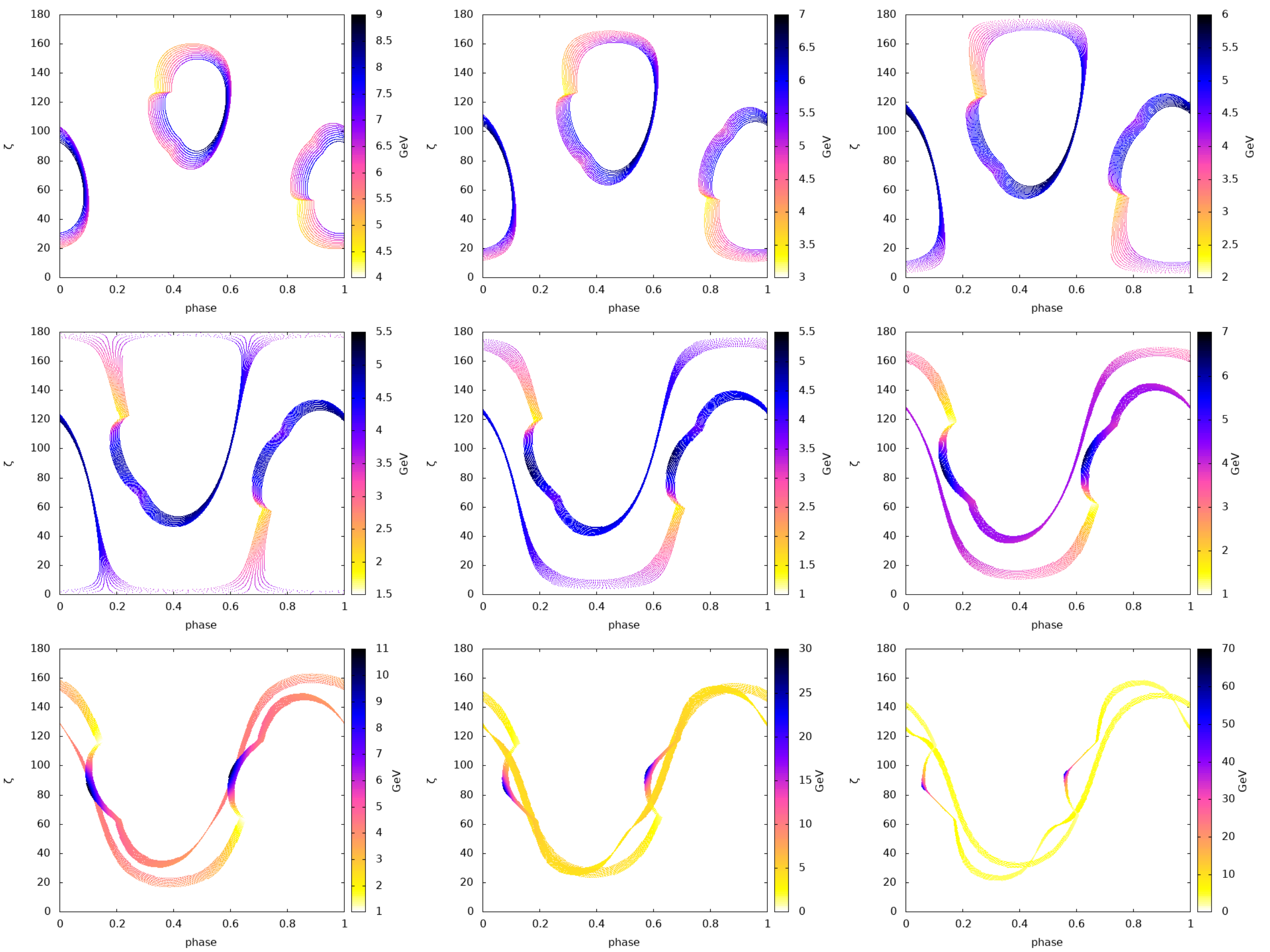}
	\caption{\label{alt60}Same as Fig.~\ref{alt30} but with an inclination $\rchi=60\degree$.}
\end{figure}

\begin{figure}
	%\makebox[\columnwidth][c]{
	\includegraphics[width=\columnwidth]{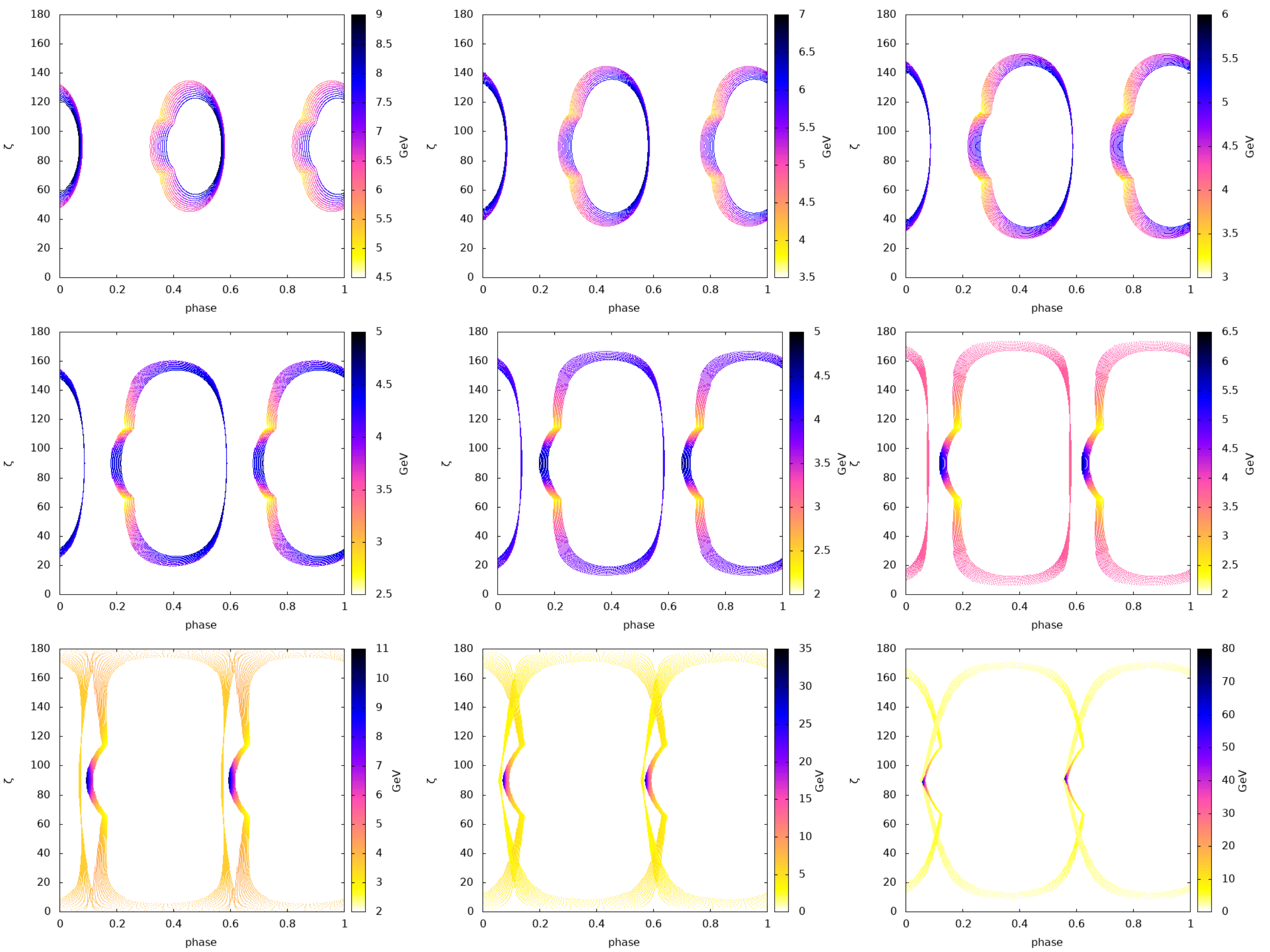}
	\caption{\label{alt90}Same as Fig.~\ref{alt30} but with an inclination $\rchi=90\degree$.}
\end{figure}

Since the frequency of the curvature radiation emitted within the slot gaps has been calculated and since the position of the emission points along the last field lines is known, we can deduce from which altitude in the magnetosphere the energy of the radiation originates. Figs.~\ref{alt30},~\ref{alt60} and \ref{alt90} show this distribution for an infinitely thin emission zone and for different inclinations of the magnetic axis, respectively of~$\rchi=30\degree$, $60\degree$ and $90\degree$. Each of these maps represents the points of impact on the celestial sphere of photons emitted between two different altitudes separated by a distance equivalent to one times the radius of the star~$R_{\star}$. The altitude of this spherical corona is plotted from $R_{\star}$ to the light cylinder of radius $10\,R_{\star}$. The colour code of these figures reflects the energy observed for the photons at the origin of each point of impact. Thus each of these three figures is presented with the arrangement shown in Table~\ref{tab:arrangement} for the range of the position~$r$ of the photon emission point.
\begin{table}[h]	
{\renewcommand{\arraystretch}{2}
\begin{center}
	\begin{tabular}{|c|c|c|}
		\hline
		$[R_{\star},2\,R_{\star}]$ & $[2\,R_{\star},3\,R_{\star}]$ & $[3\,R_{\star},4\,R_{\star}]$ \\
		\hline
		$[4\,R_{\star},5\,R_{\star}]$ & $[5\,R_{\star},6\,R_{\star}]$ & $[6\,R_{\star},7\,R_{\star}]$ \\
		\hline
		$[7\,R_{\star},8\,R_{\star}]$ & $[8\,R_{\star},9\,R_{\star}]$ & $[9\,R_{\star},10\,R_{\star}]$ \\
		\hline
	\end{tabular}
\end{center}
}
\caption{\label{tab:arrangement}Arrangement of the figures \ref{alt30}, \ref{alt60} and \ref{alt90} showing the emission altitude for each panel.}
\end{table}

These plots emphasise that the most energetic photons are emitted far away from the star surface and therefore close to the light cylinder, which could also indicate that the last closed magnetic field lines are more curved at larger distances from the neutron star. %The two high-energy emission regions encountered in the previous section in Fig.~\ref{fig:carte_HE_31,6-100GeV} are the two high-energy emission regions which represent the distribution of emission points in the light cylinder with the energy of the emitted radiation. 
It shows that these points are located in the most remote parts of the magnetosphere and that their high energy may be due to the magnetic field being dragged by the rotation of the neutron star which generates a large curvature of the magnetic field lines. 
The large curvature of the magnetic field lines responsible for the high-energy photons is shown in Fig.~\ref{img:carte_courb} where the radius of curvature of the field lines calculated at each emission point is shown (the smaller the radius of curvature, the greater the curvature). However, only the high-energy emission points have a small radius of curvature, if the higher energy emission areas are located in the upper part of the magnetosphere, this may also be due to the Doppler effect caused by the rotation of the pulsar. The Doppler factor~$\eta$ given by equation~\eqref{eq:facteur_doppler} is indeed larger for emission points located at a high altitude from the neutron star because their linear velocity $\vec{\beta}$ is larger. Fig.~\ref{img:carte_doppler} gives a map of the emission points with the Doppler factor calculated for each of them, this factor is more important near the high energy emission points.
High energy emission near the light cylinder therefore requires a favourable combination of several effects: curvature of the magnetic field lines and blue shifting Doppler factor.

\section{Radio emission\label{sec:radio}}

As is well known, the radio emission in the polar caps is caused by the acceleration of secondary pairs, that is particles generated by magnetic photon absorption in the strong magnetic field of the pulsar \citep{erber_high-energy_1966}, producing an avalanche of leptons. The Lorentz factor of these secondary particles is of the order of $\gamma=10^{2}$, as stated in \cite{beskin_physics_1993}. In what follows, we choose a mono-energetic distribution function with Lorentz factor~$\gamma=30$ in order to obtain frequencies calculated with equation~\eqref{eq:power} in the observational window of radio pulsars.

In order to avoid a sharp on/off boundary within the polar cap, not reflecting the real radio profiles, a weight is assigned to each photon emitted from the polar cap according to the angular distance~$\theta$ of these points from the magnetic pole, so that the emission profile resembles a Gaussian function. The technique is similar to the introduction of a thickness in the slot gaps described in the previous section. Thus, when plotting the emission maps and the associated light curves, the intensity received over a $0.5\times0.5$ degree zone of the celestial sphere will not increase by one unit for each photon received. Instead it increases by a weighted intensity~$I$ taking into account the Gaussian shape for photons. Photons coming from the north polar cap are weighted according to
\begin{equation}
I_{\rm north}=e^{-\theta^{2}/(\sigma \, \theta_{\rm pc})^{2}}
\label{eq:gauss_radio_nord}
\end{equation}
and a similar expression holds for the south polar cap. The parameter $\sigma = 1/\sqrt{10}$ controls the width of the Gaussian shape, the angles $\theta$ and $\theta_{\rm pc}$ being respectively the colatitude of the emission point and of the rim of the polar cap in a coordinated system oriented according to the magnetic axis. The Gaussian factor seen in equation~\eqref{eq:gauss_radio_nord} is used in order to keep pulses of Gaussian shapes and maximum emission at the centre of the polar cap. 

\begin{figure}
	\includegraphics[width=\columnwidth]{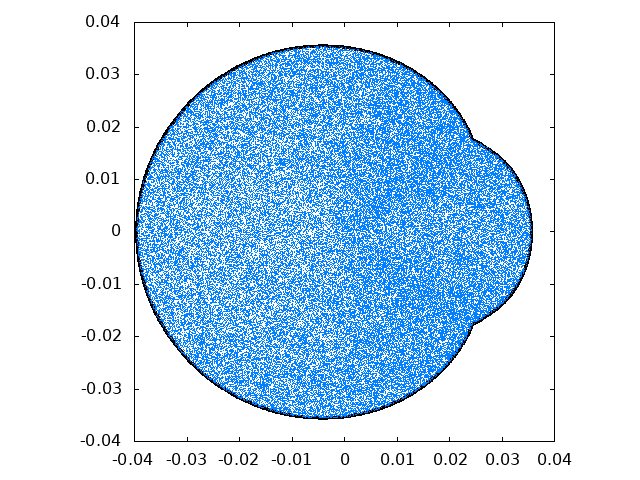} 
	\caption{\label{fig:echantillon}Distribution of $10^8$ radio emission points inside the "North" polar cap at the surface of the neutron star for a perpendicular rotator showing an almost uniform distribution.}
\end{figure}

We generated a random but homogeneous distribution of $10^8$~points inside both polar caps as depicted in Fig.~\ref{fig:echantillon}. To obtain such a distribution, the colatitude $\theta$ of the emission points in the spherical coordinate system oriented along the magnetic axis (where the colatitude~$\theta_{\rm mp}$ of the north and south pole are respectively $0\degree$ and $180\degree$) is determined through the formula
\begin{equation}
\theta - \theta_{\rm mp} = \arccos \left[ 1 - \left( 1 - \cos\theta_{\rm pc} \right) \, X \right]
\end{equation}
with $\theta_{\rm pc}$ the latitude of the point where the last closed field line crosses the stellar surface and $X$ is a random number in $[0,1]$. The longitude~$\varphi$ of the emission point is determined by introducing another random number~$Y$ in $[0,1]$ such that
\begin{equation}
\varphi = \varphi_i + Y \, \Delta \varphi
\end{equation}
where $\Delta \varphi = 2\,\pi/N_\varphi$ is the polar angle between two neighbouring points of a polar cap rim and $N_\varphi$ the number of point samples in the longitudinal direction.

\subsection{Sky maps}

\begin{figure}
	\includegraphics[width=\columnwidth]{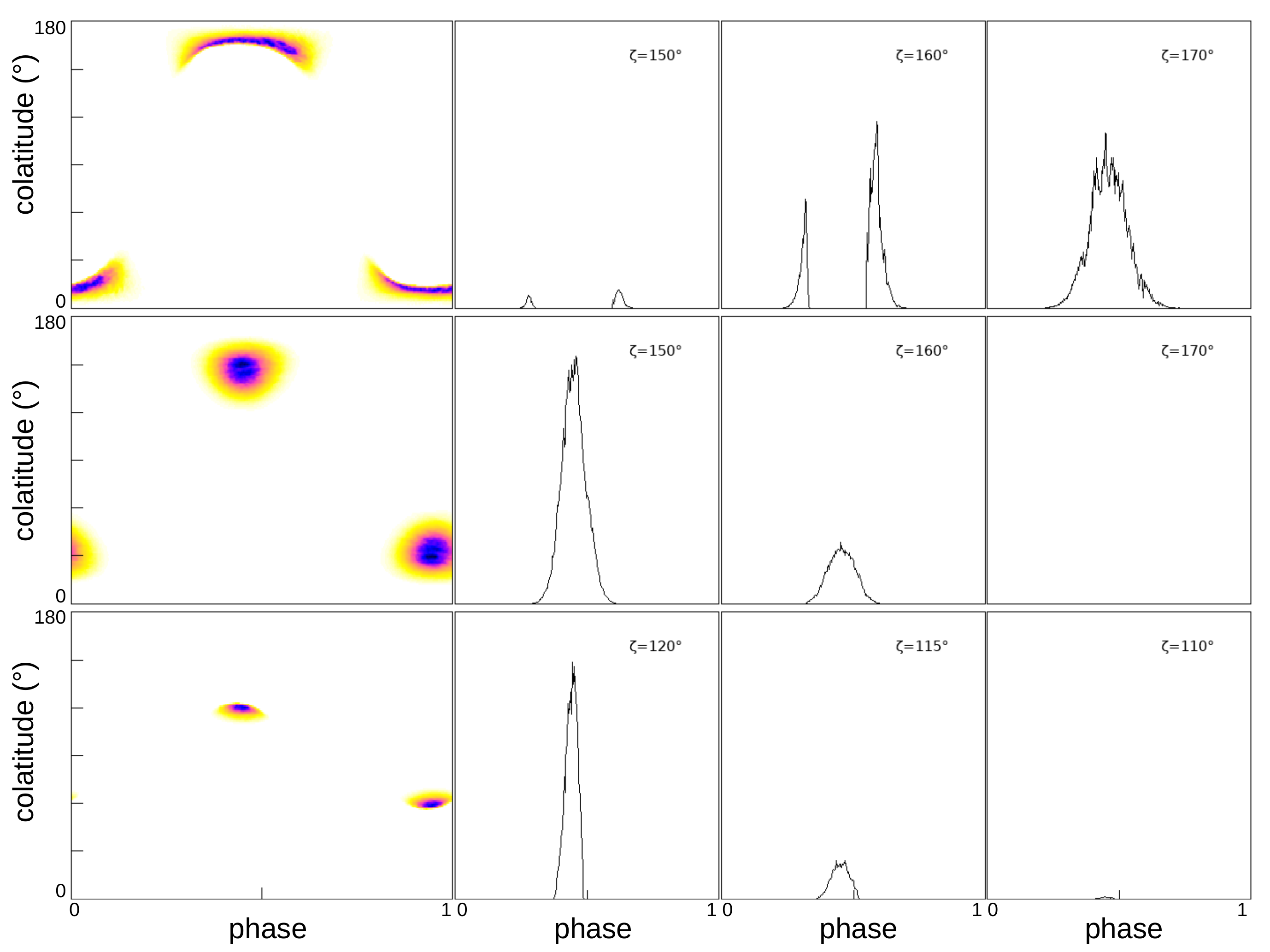} 
	\caption{\label{carte_radio_freq_30_1-2}Radio sky maps and some light curves for an emission emanating from an altitude $r\in[2,3]~R_{\star}$ for an obliquity $\rchi=30\degree$. From top to bottom, these maps shows emission frequencies within $[30,120]$~MHz, $[120,210]$~MHz and $[210,300]$~MHz.}
\end{figure}

\begin{figure}
	\includegraphics[width=\columnwidth]{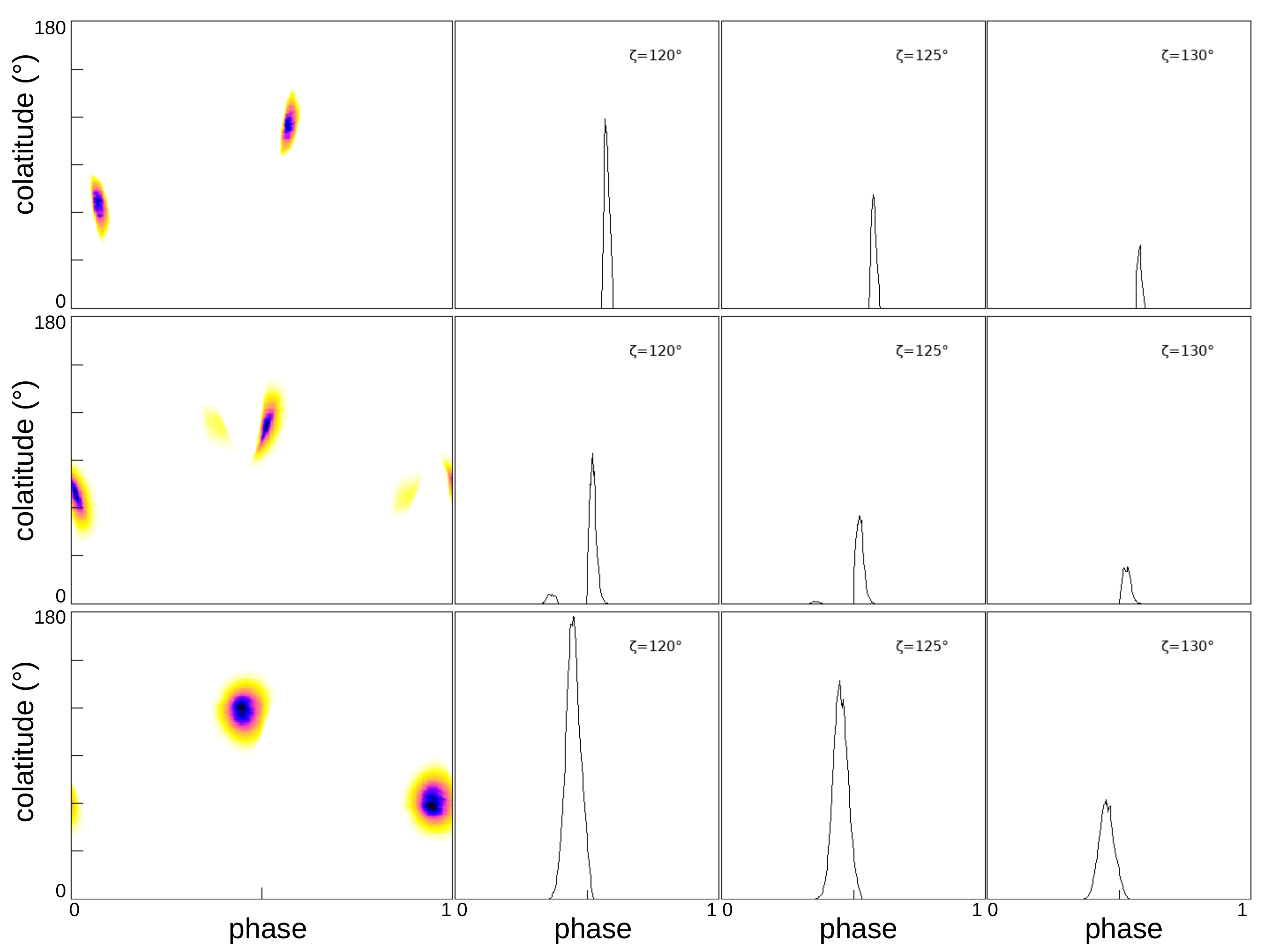} 
	\caption{\label{carte_radio_freq_60_1-2}Radio sky maps and some light curves for an emission emanating from an altitude $r\in[2,3]~R_{\star}$ for an obliquity $\rchi=60\degree$. From top to bottom, these maps shows emission frequencies within $[200,270]$~MHz, $[270,340]$~MHz and $[340,410]$~MHz.}
\end{figure}

\begin{figure}
	\includegraphics[width=\columnwidth]{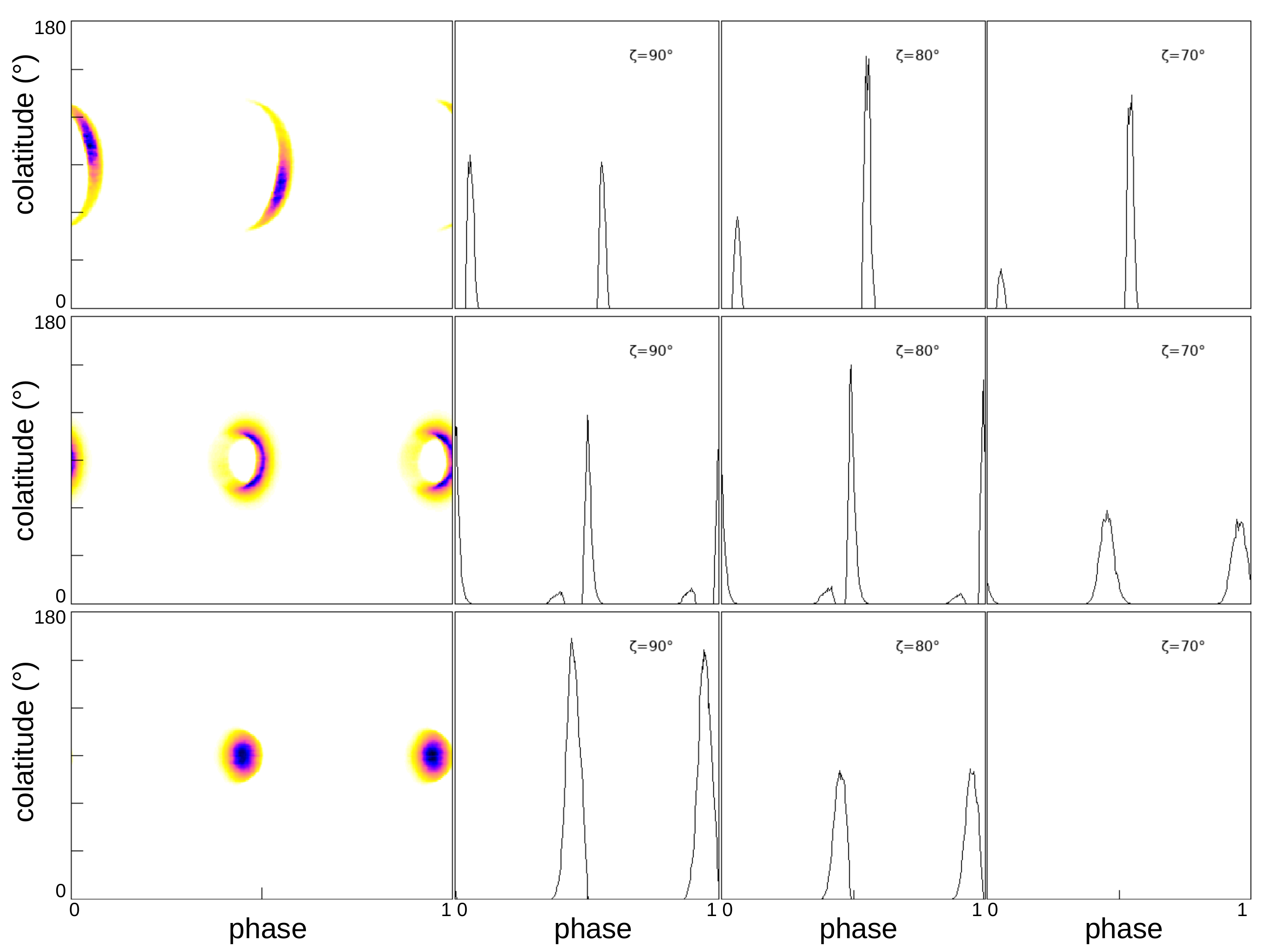} 
	\caption{\label{carte_radio_freq_90_1-2}Radio sky maps and some light curves for an emission emanating from an altitude $r\in[2,3]~R_{\star}$ for an obliquity $\rchi=90\degree$.  From top to bottom, these maps shows emission frequencies within $[330,450]$~MHz, $[450,570]$~MHz and $[570,690]$~MHz.}
\end{figure}

Let us now determine the radio emission maps as a function of frequency in order to study the evolution of the pulse shape as measured by a distant observer. Figs.~\ref{carte_radio_freq_30_1-2}, \ref{carte_radio_freq_60_1-2} and \ref{carte_radio_freq_90_1-2} show examples of sky maps and a corresponding sample of light curves obtained for an emission volume in the range $r\in[2,3]~R_{\star}$ above the polar caps for different frequency intervals. Note that the frequency interval breakdown is not the same for the different obliquities~$\rchi$ because the spectra do not have the same range or the same upper and lower limits. The choice of frequencies has been optimized to make the intensity maps stand out best. 

These maps reveal a narrowing of the pulse profile towards the high frequencies since these photons come from the deepest regions of the magnetosphere. There is also an east-west asymmetry in these profiles between the rising and falling ramps. Sometimes two pulses are observed in the profile, sometimes only one. It should be stressed that these results are only preliminary and that any comparison with observations will have to take into account a particle energy distribution in the form of a power law and not simply a mono-energetic distribution as done in the present work, see the particle distribution function discussion in section~\ref{sec:Model}. The characteristic curvature frequency must also be replaced by the continuous curvature spectrum centred on this typical frequency. This will make the frequency variation of the pulse profile smoother and the transition between the frequency bands more continuous. After this further step, however left as a future work, detailed comparison with existing multi-wavelength observations can be performed.

\begin{figure*}
	\centering
	\begin{tabular}{cc}
		\includegraphics[width=0.5\columnwidth]{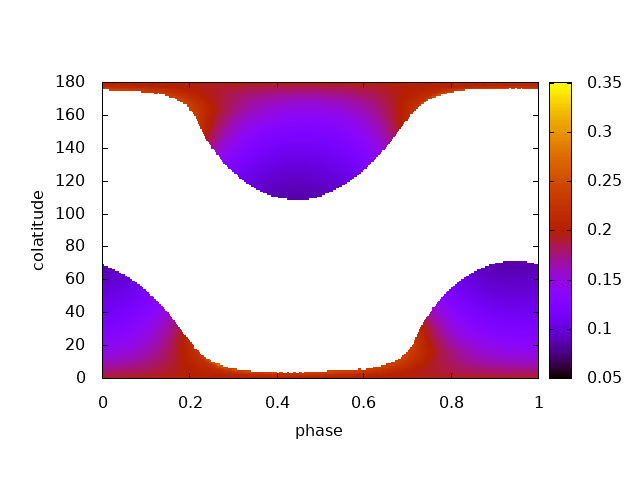} &
		\includegraphics[width=0.5\columnwidth]{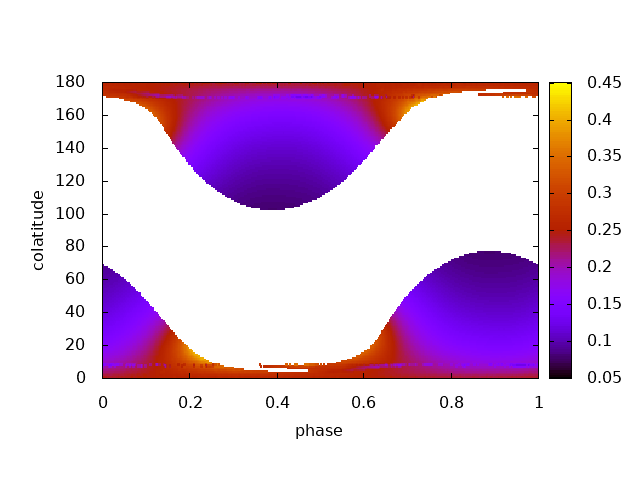}
	\end{tabular}
	\caption{\label{carte_courb_radio30}Maps of curvature radii in units of $R_{\rm cyl}$ for an inclination $\rchi=30\degree$, on the left for an altitude in $[2,3]~R_{\star}$, on the right for $[4,5]~R_{\star}$ above the polar caps.} 
\end{figure*}

\begin{figure*}
	\centering
	\begin{tabular}{cc}
		\includegraphics[width=0.5\columnwidth]{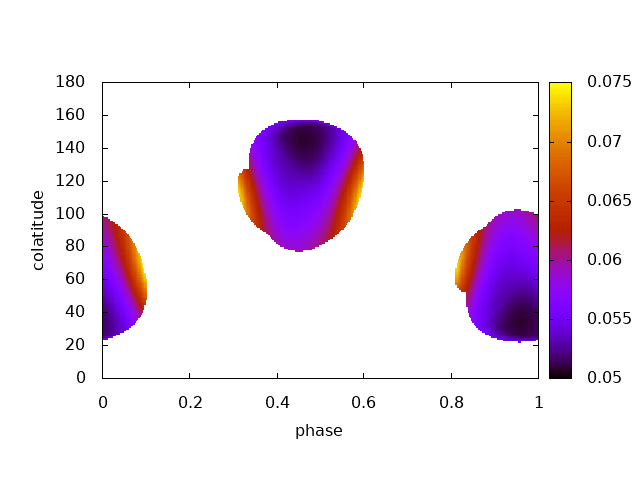} &
		\includegraphics[width=0.5\columnwidth]{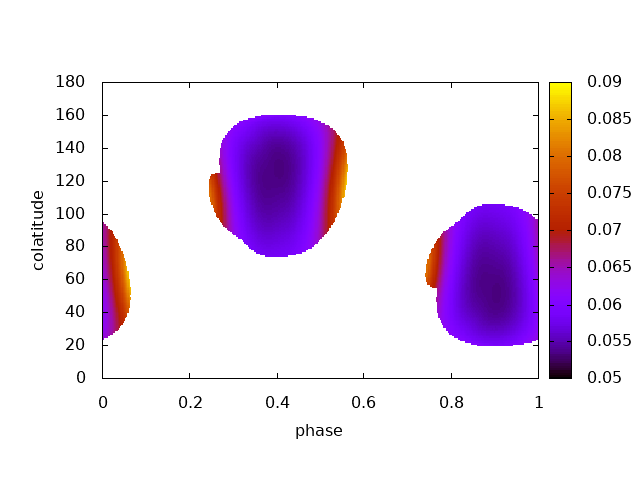}
	\end{tabular}
	\caption{\label{carte_courb_radio60}Same as Fig.~\ref{carte_courb_radio30} fut for an inclination $\rchi=60\degree$.} 
\end{figure*}

\begin{figure*}
	\centering
	\begin{tabular}{cc}
		\includegraphics[width=0.5\columnwidth]{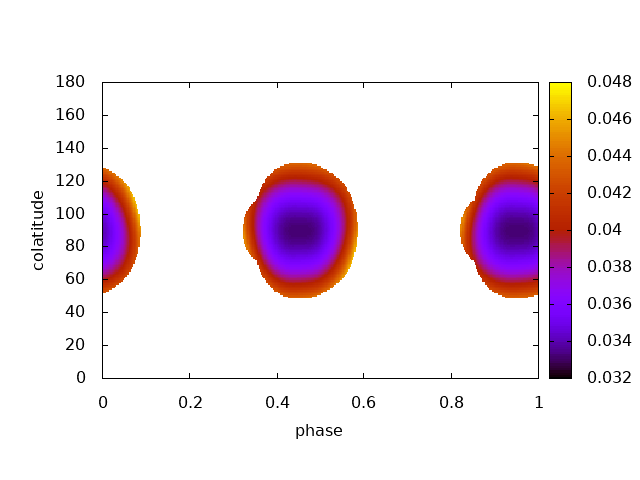} &
		\includegraphics[width=0.5\columnwidth]{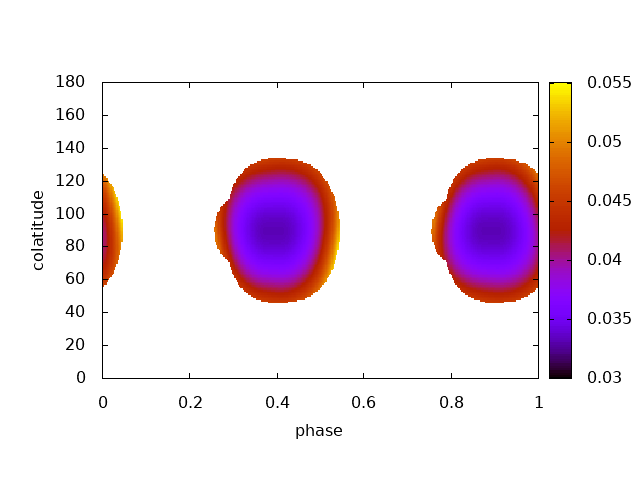}
	\end{tabular}
	\caption{\label{carte_courb_radio90}Same as Fig.~\ref{carte_courb_radio30} fut for an inclination $\rchi=90\degree$.} 
\end{figure*}

According to these maps, especially those in Fig.~\ref{carte_radio_freq_90_1-2}, low frequencies are emitted rather at the edge of the area while the higher frequency components are emitted towards the centre of the area. This may seem counter-intuitive since for a static dipole, the centre of the pulse, corresponding to the magnetic axis, is a straight line of zero curvature and therefore of zero frequency and power. In other words, within the static limit, there is no emission at the centre of the cap. Frequencies and power increase with increasing angular distance from the axis. But for a rotating dipole, as for example in the Deutsch solution we use in the present study, the magnetic field undergoes a sweep-back effect curving all field lines, even the one associated with the magnetic axis. To be able to visualise this curvature, figs.~\ref{carte_courb_radio30}, \ref{carte_courb_radio60} and \ref{carte_courb_radio90} represent maps of the radii of curvature, measured in units of $R_{\rm cyl}$, at the point of photon emission for $\chi=30\degree,60\degree,90\degree$ respectively and for two emission heights, on the left for $[2,3]~R_\ast$ and on the right for $[4,5]~R_\ast$. 
%Only photon with the value of the curvature of the maximum field line (and thus the smallest radius of curvature) at its emission point is displayed on each area of $0.5\degree$ over $0.5\degree$. Two altitudes are shown, left for $[1.2]~R_{\star}$ and right for $[3.4]~R_{\star}$. 
The fact that the highest frequencies are towards the centre of the polar caps is therefore a consequence of the magnetic field line curvature induced by the stellar rotation. The radius of curvature on the field lines from the polar caps is minimal in the case of the perpendicular rotator with $\rchi=90\degree$ and reaches $0.032\,R_{\rm cyl}$. It increases when the magnetic axis joins the rotation axis, i.e. when $\rchi$ decreases. For example, for $\rchi=30\degree$ this minimum radius is $0.05\,R_{\rm cyl}$ and moves outside the centre of the cap. At the same time the maximum curvature radius increases significantly when the rotator is close to the aligned configuration. It is only $0.048\,R_{\rm cyl}$ for $\rchi=90\degree$ and reaches up to $0.35\,R_{\rm cyl}$ for $\rchi=30\degree$ which is an increase of almost one order of magnitude. The photons produced will be much less energetic in this latter case. Within the limit of a perfectly aligned rotator $\rchi=0\degree$, the radius of curvature becomes infinite at the centre of the cap, on the magnetic axis, and the curvature radiation is extinguished at its centre.


\begin{figure}
	\includegraphics[width=\columnwidth]{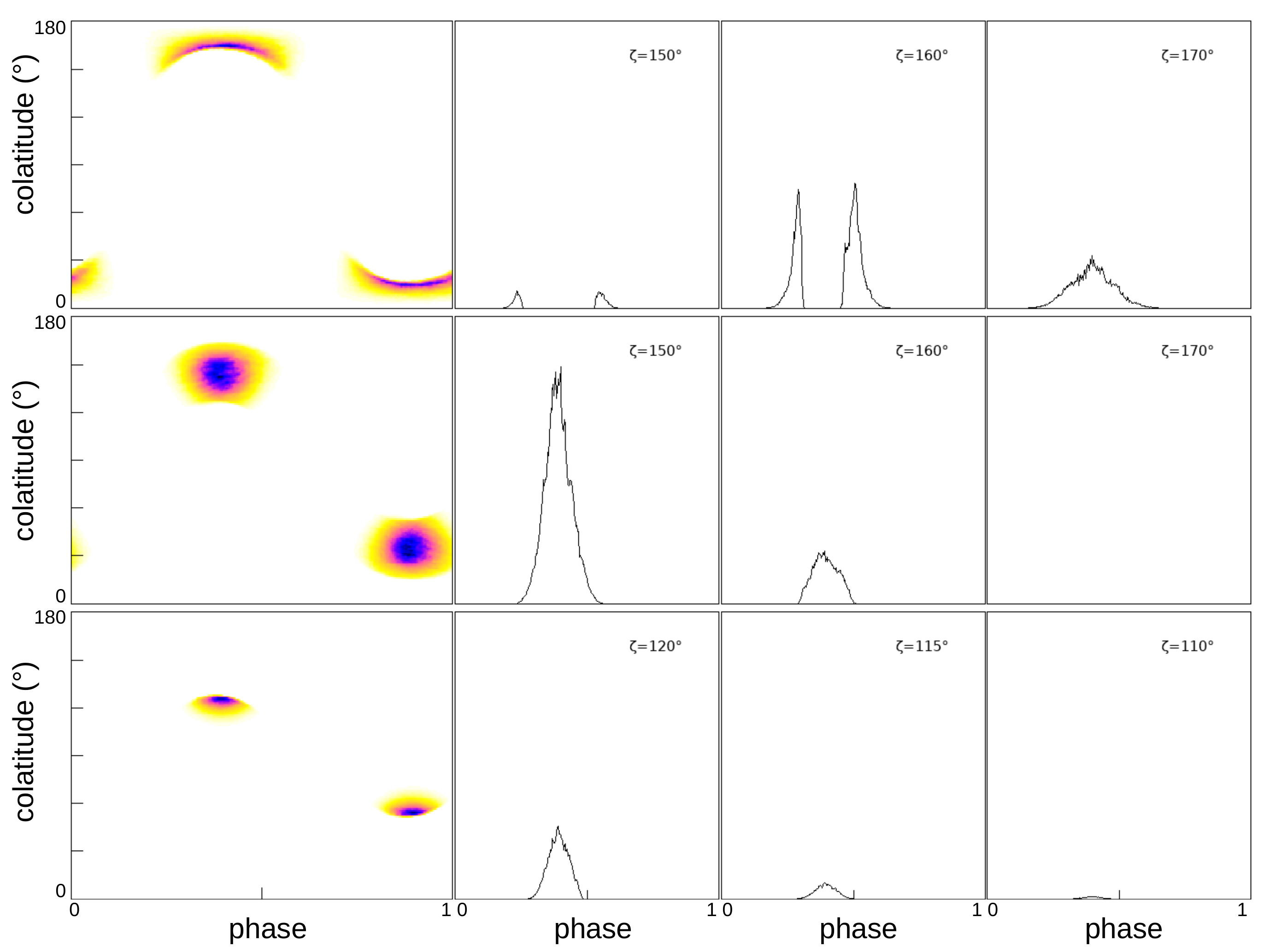} 
	\caption{\label{carte_radio_freq_30_3-4}Radio sky maps and some light curves for an emission emanating from an altitude $r\in[4,5]~R_{\star}$ for an obliquity $\rchi=30\degree$. From top to bottom, these maps shows emission frequencies within $[30,120]$~MHz, $[120,210]$~MHz and $[210,300]$~MHz.}
\end{figure}

\begin{figure}
	\includegraphics[width=\columnwidth]{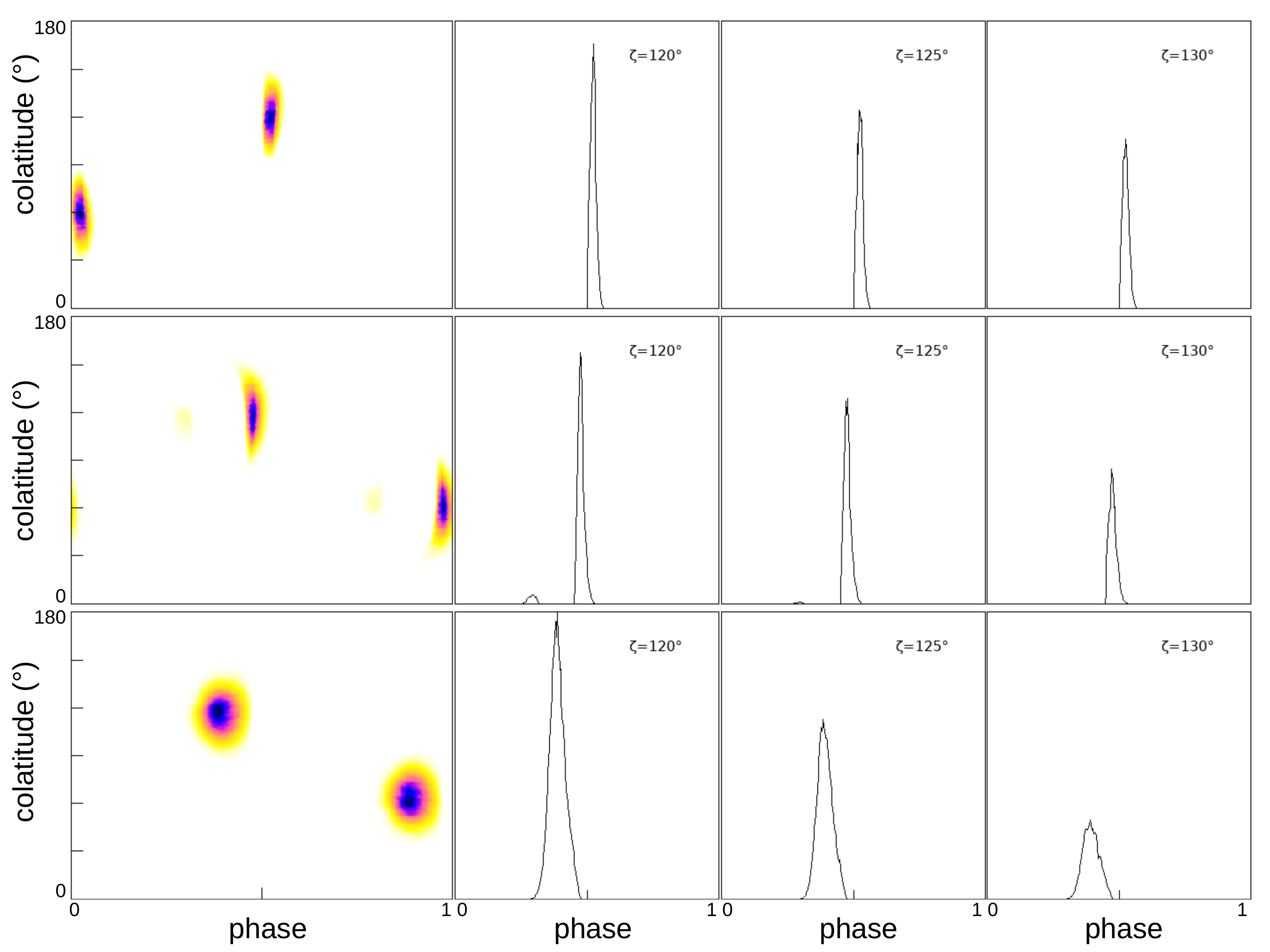} 
	\caption{\label{carte_radio_freq_60_3-4}Radio sky maps and some light curves for an emission emanating from an altitude $r\in[4,5]~R_{\star}$ for an obliquity $\rchi=60\degree$. From top to bottom, these maps shows emission frequencies within $[200,270]$~MHz, $[270,340]$~MHz and $[340,410]$~MHz.}
\end{figure}

\begin{figure}
	\includegraphics[width=\columnwidth]{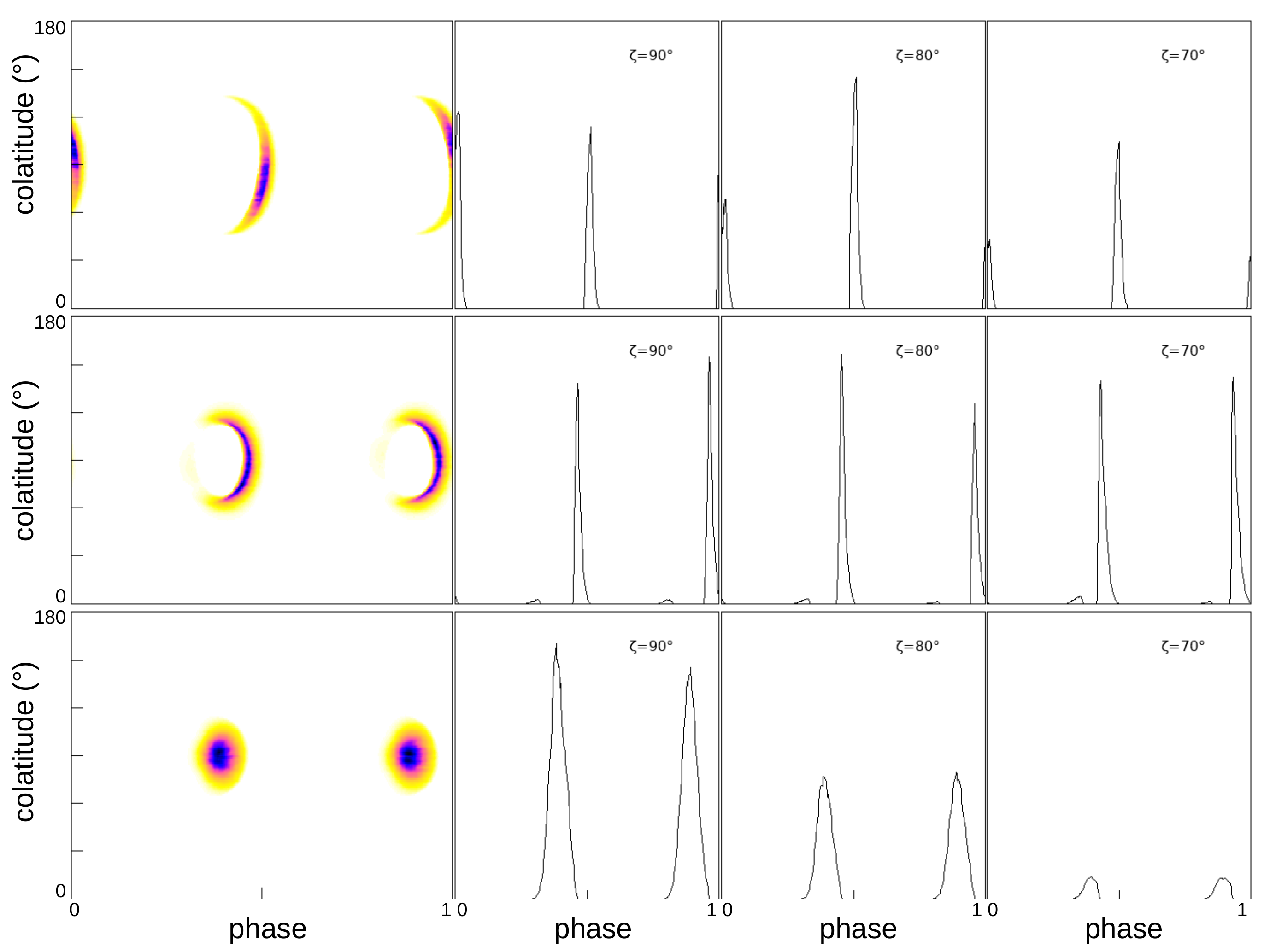} 
	\caption{\label{carte_radio_freq_90_3-4}Radio sky maps and some light curves for an emission emanating from an altitude $r\in[4,5]~R_{\star}$ for an obliquity $\rchi=90\degree$.  From top to bottom, these maps shows emission frequencies within $[330,450]$~MHz, $[450,570]$~MHz and $[570,690]$~MHz.}
\end{figure}

The altitude at which photons are produced and their detachment from the magnetosphere in the direction of the observer is not precisely constrained by the observations. The radio polarisation data provide some indication for young pulsars, but the error bar remains appreciable. There is therefore some freedom in the choice of the height of the emission sites above the polar ice caps. For this reason we have also drawn emission maps and light curves for an altitude in the range $[3,4]~R_{\star}$ as shown in Figs.~\ref{carte_radio_freq_30_3-4}, \ref{carte_radio_freq_60_3-4} and \ref{carte_radio_freq_90_3-4} for the same frequency intervals above the polar caps.

Pushing the emission altitude further away from the star causes a spreading of the radio pulse profile since the field lines are divergent. This is reflected in the fact that high altitudes produce the lowest frequencies in relation to wider pulses while low altitudes produce the highest frequencies in relation to narrower pulses, in accordance with the radius-to-frequency mapping model. But in our case, we dispense with the static dipole to take into account all the effects due to the rotation of the magnetic field, the effects of signal aberration and delay.


\begin{figure}
	\centering
	\includegraphics[width=\columnwidth]{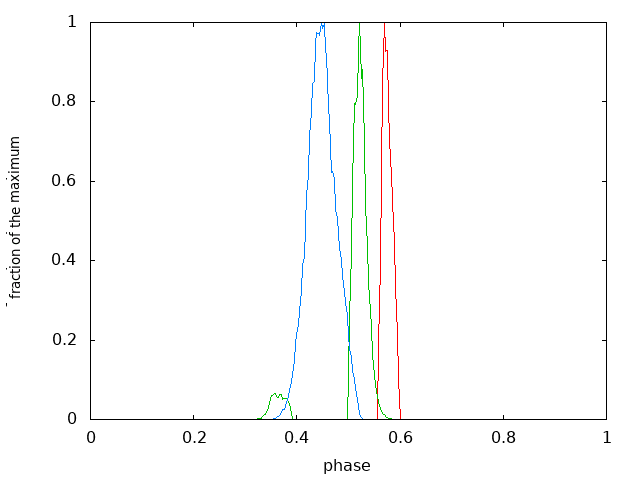} 
	\caption{\label{lc_radio_60-120_1-2}Radio light-curves for $\chi=60\degree$, $\zeta=120\degree$ and an emission zone  $r\in[2,3]~R_{\star}$. In red for frequencies in the interval $[200,270]$~MHz, in green for $[270,340]$~MHz and in blue for $[340,410]$~MHz.}
\end{figure}

In order to facilitate the visualisation of the evolution of the radio pulses as a function of frequency, the shape of a light curve has been plotted in Fig.~\ref{lc_radio_60-120_1-2} according to the frequency interval considered, here those obtained with an emission area between $r\in[2,3]~R_{\star}$ above the polar caps and for an inclination~$\rchi=60\degree$ of the magnetic axis and~$\zeta=120\degree$ of the line of sight. The maximum of the blue profile in the range [340,410]~MHz is ahead of the green profile in the range [270,340]~MHz, which in turn is ahead of the red pulse in the range [200,270]~MHz. This phase lead is significant since it represents 5 to 10\% of the pulsar period. It is the consequence of the curvature of the magnetic field lines in the retrograde direction with respect to the stellar rotation and of the production of the high-frequency photons at low altitude, where the field lines are less curved compared to the low-frequency photons produced at higher altitudes where the field lines are more curved, due to the sweep back effect, eq.\eqref{eq:retard_shitov}. This effect depends on the ratio $(R_{\ast}/R_{\rm cyl})$, it is negligible for young pulsars but its observational signature should be perceptible for the fastest millisecond pulsars.

The phase shift can be estimated from eq.\eqref{eq:decal_ret_mink}, \eqref{eq:decal_aberr} and \eqref{eq:retard_shitov} in paragraph~\ref{sec:retard}. The maximum offset thus calculated (for a photon emitted at $1~R_{\star}$ above the surface and another emitted at $2~R_{\star}$ and always with $\chi=60\degree$) is of the order of 4\% of the phase. These effects are therefore not sufficient to explain the shift of the radio emission peak in Fig.~\ref{lc_radio_60-120_1-2} but they can certainly play a role in the radio emission observed in these emission areas.

\subsection{Spectra}

\begin{figure}
	\centering
	\includegraphics[width=\columnwidth]{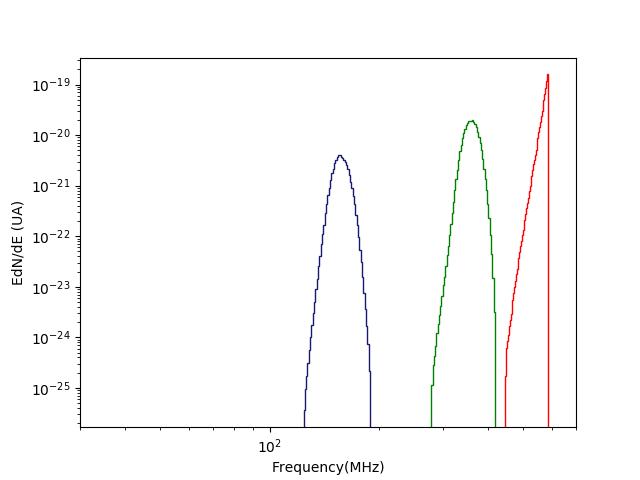}
	\caption{\label{Spectre_radio_surface}Radio emission spectrum, in blue for $\rchi=30^{\circ}$, in green for $\rchi=60^{\circ}$ and in red for $\rchi=90^{\circ}$. } 
\end{figure}

As for the high energy radiation, fig.~\ref{Spectre_radio_surface} shows the radio spectrum for different values of the obliquity~$\rchi$ of the magnetic field. The points of emission for this spectrum are located on the stellar surface and with the same spatial distribution as shown in fig.~\ref{fig:echantillon}.

Fig.~\ref{Spectre_radio_surface} reveals that the frequency range of the radio emission depends strongly on the angle~$\rchi$, because the curvature of magnetic field lines is more pronounced for the perpendicular rotator $\rchi=90\degree$ compared to a nearly aligned rotator $\rchi=30\degree$. Indeed, it is for a magnetic axis perpendicular to the rotation axis that the radio frequencies are the highest, and since the magnetic polar caps are in the equatorial plane, the magnetic field lines passing through them are strongly swept back, and therefore curved by the rotation of the pulsar. % (as can be seen in Fig.~\ref{champ_mag}). 
The magnetic field lines are curved backwards, in the opposite direction of the rotation as shown by \cite{shitov_period_1983}, delaying the radio signal by an amount~$\Delta t_B$ given by \eqref{eq:retard_shitov}. The Doppler effect also contributes partly since the corotation speed $v_{\rm cp} = r\,\Omega\,\sin\rchi$ is more significant for the perpendicular rotator because of the longer lever arm of length $r\,\sin\rchi$. The amplitude of this Doppler effect is related to $\beta_{\rm cp} = (r/R_{\rm cyl})\,\sin\rchi$. Generally speaking, the characteristic radio frequency~$\nu_{\rm radio}$ depends on the ratio $(r/R_{\rm cyl})$ and on $\sin\rchi$. Therefore the range of radio waves around~$\nu_{\rm radio}$ behaves as an increasing function of $\sin\rchi$. On the other hand, this shift of the radio spectrum towards higher frequencies is attenuated for rotators slower than those studied within this paper. Indeed, for young pulsars, of period~$P$ greater than about $P>100$~ms, the ratio between the neutron star radius and the light cylinder radius is much smaller than one, $(r/R_{\rm cyl})\ll1$, hence producing a less perceptible variation in frequency as a function of $\sin\rchi$.

In the above proposed image, the photons escape directly from the star's surface. In our simulations, the period of rotation of the pulsar is limited to 2~ms because of the prohibitive computing time for a star of larger period. The neutron star surface is fixed at a radius $0.1\,R_{\rm cyl}$. Nevertheless, for slow pulsars, this radius of $0.1\,R_{\rm cyl}$ corresponds to the average real altitude of radio emission (it is in fact a little lower, of the order of $0.05\,R_{\rm cyl}$, see for instance \cite{mitra_nature_2017}). The width of the pulses as well as the delay between the radio peak and the first gamma-ray peak thus remains realistic in spite of the too high period. However, the assumption that radio emission is produced at a fixed altitude is not entirely correct. Indeed, it is known that the highest frequencies are generated at lower altitudes because of the radio-to-frequency mapping. It will therefore be necessary to include an additional degree of freedom for the true position of the emission sites by introducing, for example, an interval of variable altitudes as was done in the high energy band.

\begin{figure*}
	\centering
	\begin{tabular}{cc}
		\includegraphics[width=0.5\columnwidth]{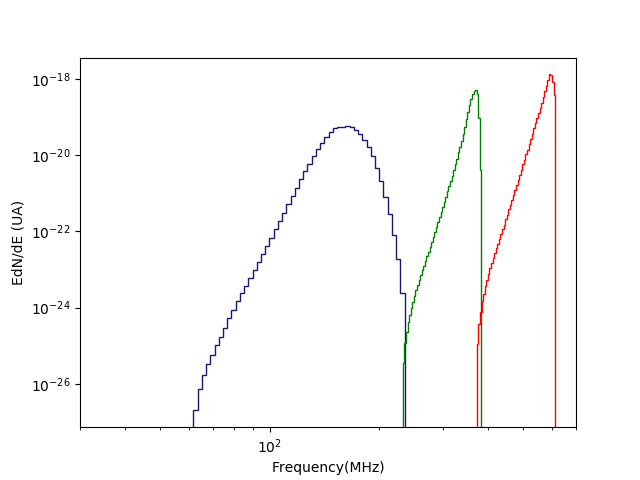} &
		\includegraphics[width=0.5\columnwidth]{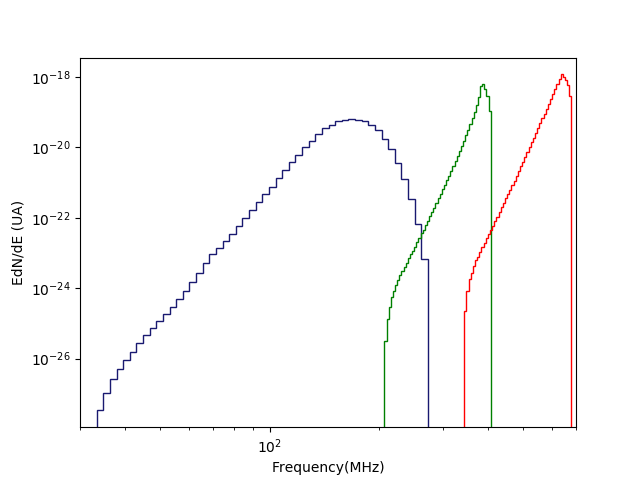}
	\end{tabular}
	\caption{\label{Spectre_radio_alt}Radio emission spectra for a region located in the range $[1,2]~R_{\star}$ on the left, and in the range $[3,4]~R_{\star}$ on the right, in blue for $\rchi=30^{\circ}$, in green for $\rchi=60^{\circ}$ and in red for $\rchi=90^{\circ}$.} 
\end{figure*}

The figures in~\ref{Spectre_radio_alt} show the radio spectrum for different values of the obliquity~$\rchi$, when the emission area is located above the polar caps, respectively in the range $r\in[1,2]~R_{\star}$ on the left part and in the range $r\in[3,4]~R_{\star}$ on the right part. Compared to Fig.~\ref{Spectre_radio_surface}, the emission no longer occurs in a given radius~$r$ but for an entire range of radii~$r\in[r_1,r_2]$. The emission is no longer integrated on a surface (the polar cap) but in a whole three-dimensional volume. The result is a wider range of values of radii of curvature and consequently a wider radio emission spectrum, regardless of the value of~$\rchi$. This spread is maximum for $\rchi=30\degree$ and less for other values of the inclination of the magnetic axis such as $60\degree$ and $90\degree$. Shifting the emission volume to a higher altitude produces an additional spread of the spectrum, which can be seen by comparing the left and right plots in fig.~\ref{Spectre_radio_alt}.
We conclude from these spectra that the farther the radio emission zone from the surface of the neutron star, the wider the frequency range of the received radio radiation. This phenomenon is very important for the smallest inclination investigated, namely for $\rchi=30\degree$. 

This broadening of the spectra is due to the fact that our emission zones above the polar caps possess a certain thickness and is therefore the result of the variation in the curvature of magnetic field lines. However, other phenomena may affect the shape of these spectra, such as the geometry of the field lines, which is certainly varying with altitude, or also the Doppler factor, which depends on the instantaneous rotation speed $\vec{v} = \Omega\wedge\vec{r} = r\,\Omega\,\sin\theta\,\vec{e}_\varphi$ and increases with altitude due to the corotation of the magnetosphere with the neutron star.

\section{Conclusion\label{sec:Conclusion}}

We studied one of the main emission mechanisms taking place within the pulsar magnetosphere, namely the curvature radiation from the slot gap and polar cap inside the light cylinder. We showed detailed radio and high-energy light-curves and spectra in the Fermi/LAT band. Using realistic values of the Lorentz factors for primary and secondary beams, our spectra fall into the right windows. We computed phase-resolved spectra and found that the spectral energy distribution depends on the phase interval. Also, the averaged radio spectra significantly depend on the pulsar obliquity whereas the high-energy part seemed much less sensitive to the geometry.

In order to get a more complete view of this emission, it might be useful to add synchrotron radiation and inverse Compton scattering to future simulations, as well as other models with different emission zones, such as outer gaps or the striped wind. The calculation of the spectra will also gain in realism if the mono-energetic particle distribution is replaced by a power law distribution and the $\delta$ approximation of the curvature spectrum by its spectral energy distribution.

The polarisation of the pulsed emission could also be added to our simulations of radio and high energy emission. In the radio domain, this polarisation already makes it possible to shift the emission sites at high altitude for slow pulsars with periods above 100~ms. The imminent observation of the polarisation in X-rays will bring a new strong constraint on the high energy part of the spectrum, determining its location either inside the light cylinder or outside within the wind.

Finally, a discussion of these results in relation to pulsar multi-wavelength observations should constrain many dynamical parameters of the relativistic plasma by comparing, for example, the light curves as well as the radio and gamma-ray spectra for pulsars detected simultaneously in these two energy bands.

\section*{Acknowledgements}

This work has been partly supported by CEFIPRA grant IFC/F5904-B/2018. We also acknowledge the High Performance Computing center of the University of Strasbourg for supporting this work by providing scientific support and access to computing resources. Part of the computing resources were funded by the Equipex Equip@Meso project (Programme Investissements d'Avenir) and the CPER Alsacalcul/Big Data.

%\bibliographystyle{aa}
%\bibliography{/home/petri/zotero/Ma_bibliotheque}

\end{document}